\documentclass[journal]{IEEEtran}
\usepackage{amsfonts}
\usepackage{amssymb}
\usepackage{eurosym}
\usepackage{cite}
\usepackage{graphicx}
\usepackage{epstopdf}
\usepackage{amsmath}
\usepackage{tikz,lipsum}
\usepackage{caption}
\usepackage[T1]{fontenc}
\usepackage{amsthm}
\usepackage{mathrsfs}
\usepackage{color}
\usepackage{stfloats}
\usepackage{xcolor, etoolbox}
\usepackage{subcaption}

\IEEEoverridecommandlockouts
\newcounter{mytempeqncnt}
\newtheorem{remark}{Remark}

\newtheorem{lemma}{Lemma}
\newtheorem{proposition}{Proposition}

\begin{document}

\title{Can Artificial Noise Boost Further the Secrecy of Dual-hop RIS-aided
Networks?}
\author{Elmehdi Illi, \IEEEmembership{Member, IEEE}, Marwa K. Qaraqe, %
\IEEEmembership{Senior Member, IEEE}, Faissal El Bouanani, %
\IEEEmembership{Senior Member, IEEE}, and Saif M. Al-Kuwari,
\IEEEmembership{Senior Member,
IEEE} \thanks{%
This paper was submitted in part to BalkanCom 2022 \cite{balkan}.} \thanks{
This research was sponsored in part by the NATO Science for Peace and
Security Programme under grant SPS G5797.} \thanks{%
E. Illi, M. K. Qaraqe, and S. M. Al-Kuwari are with the College of Science
and Engineering, Hamad Bin Khalifa University, Qatar Foundation, Doha,
Qatar. (e-mails: elmehdi.illi@ieee.org, \{mqaraqe, smalkuwari\}@hbku.edu.qa).%
} \thanks{%
F. El Bouanani is with ENSIAS College of Engineering, Mohammed V University
of Rabat, Morocco. (e-mail: f.elbouanani@um5s.net.ma).} }
\maketitle

\begin{abstract}
In this paper, we quantify the physical layer security of a dual-hop
regenerative relaying-based wireless communication system assisted by
reconfigurable intelligent surfaces (RISs). In particular, the setup
consists of a source node communicating with a destination node via a
regenerative relay. In this setup, a RIS is installed in each hop to
increase the source-relay and relay-destination communications reliability,
where the RISs' phase shifts are subject to quantization errors. The
legitimate transmission is performed under the presence of a malicious
eavesdropper attempting to compromise the legitimate transmissions by
overhearing the broadcasted signal from the relay. To overcome this problem,
we incorporate a jammer to increase the system's secrecy by disrupting the
eavesdropper through a broadcasted jamming signal. Leveraging the
well-adopted Gamma and Exponential distributions approximations, the system's secrecy level is quantified by deriving approximate and
asymptotic expressions of the secrecy intercept probability (IP) metric in
terms of the main network parameters. The results show that the secrecy is
enhanced significantly by increasing the jamming power and/or the number of
reflective elements (REs). In particular, an IP of approximately $10^{-4}$
can be reached with $40$ REs and $10$ dB of jamming power-to-noise ratio
even when the legitimate links' average signal-to-noise ratios are $10$-dB
less than the eavesdropper's one. We show that cooperative jamming is very
helpful in strong eavesdropping scenarios with a fixed number of REs, and
the number of quantization bits does not influence the secrecy when
exceeding $3$ bits. All the analytical results are endorsed by Monte Carlo
simulations.
\end{abstract}




\IEEEpeerreviewmaketitle

\section{Introduction}

The past few decades witnessed significant efforts on designing
ultra-reliable, self-sustainable, and secure communications, pillars
representing the main targets of the 5G\ and beyond and 6G\ visions \cite%
{6G,6Gb}. In fact, information security has always been a \textit{de facto}
concern in wireless transmissions due to their broadcast nature \cite{access}%
. To date, the implementation of security mechanisms on wireless
communication systems (WCSs) has been viewed almost exclusively from higher
layers by grasping the key-based cryptographic algorithms \cite{access}.
However, while these algorithms provide the required security for devices
with sufficient transmit power and computing facilities onboard \cite{pla},
the emergence of new wireless paradigms, such as the internet of things
(IoT) and vehicular communications renders current cryptographic schemes
unsuited for such power-limited and processing-restricted technologies \cite%
{jehadsurvey}. Furthermore, the forecasted sizeable amount of data traffic
is expected to bring unprecedented privacy leakages \cite{6GPLS}.

Recently, physical layer security (PLS) has been gaining significant
attention from both academia and industry communities. Unlike the
conventional key-based cryptographic schemes, PLS establishes secure
transmissions leveraging exclusively the physical layer parameters (e.g.,
fading, channel coding, interference, etc) \cite{wyner,sustainable} in order
to provide a noisy legitimate signal copy to the eavesdropper (i.e.,
increasing its equivocation). The secrecy capacity is PLS's cornerstone, for
which higher values correspond to an improved system's security level from
the physical layer point of view. To this end, PLS\ can effectively
contribute to providing acceptable security levels with a much-reduced
overhead in comparison with the traditional cryptographic schemes \cite{pla}.

\bigskip Regular WCSs consist of a transmitter sending an
information-bearing signal to a receiver through an uncontrollable
propagation medium. This signal reaches the destination via several replicas
through multiple reflection paths, producing random fading \cite{emilris}.
However, futuristic wireless networks such as the 6G are envisioned to adopt
the \textit{Smart Radio Environment,} where every network component can
adapt to the changes in the environment \cite{emilmag}. To this end,
considerable attention has been paid in the yesteryears to the \textit{%
Reconfigurable Intelligent Surface} (RIS) technology as a key enabler to
spectrum and energy efficiency's boosting \cite{refris}. A RIS\ is a
man-made metasurface consisting of a large number of low-cost passive
reflecting elements (REs), where each of which can tune its phase shift to
adapt to the incident electromagnetic wave impinging its surface. As a
result, such reflected signal copies can be constructively/destructively
superposed at the intended/unintended node to maximize/minimize the received
signal power; a process that resembles the well-known MIMO beamforming \cite%
{ris2}.

The inherent capabilities of RISs and metasurfaces in reshaping the
propagation environment have driven intuitive insights into exploiting these
features in favor of boosting the system's secrecy. As illegitimate network
users can benefit from different signal qualities, which poses serious
eavesdropping threats, RIS can be used to effectively choose optimal phase
shifts in favor of the legitimate user; i.e., beamsteer the information
signal to the genuine user with a higher power, while providing the
eavesdropper with a lower signal power.

\subsection{Related Work}

The corresponding literature of related work involving PLS\ and RIS\ has
done a substantial contribution to the subject so far. In particular, the
authors in \cite{lit4} consider the PLS of a two-way multi-user RIS-assisted
transmission for which a user scheduling scheme was proposed. In \cite{lit1}%
, the secrecy level of a\ RIS-assisted WCS was tackled where a direct link
between the transmitter and the receiver was considered. In \cite{lit2}, the
authors carried out a PLS analysis\ of a RIS-based two-users non-orthogonal
multiple access network. Similarily, the work in \cite{litt1} inspected the
security performance of a RIS-aided network with an uncertain eavesdropper
location. Moreover, several work dealt with the secrecy maximization problem
by optimizing the RIS and system parameters. For instance, the authors in
\cite{lit3} formulated a secrecy rate-maximization problem under the
RIS-assisted system constraints and proposed an efficient algorithm for
solving the optimization problem. Similarly, an energy-efficiency secure
transmission problem for a multi-antenna source RIS-assisted multi-user
network was formulated in \cite{litt2} with the probabilistic outage
constraint. In addition, the authors in \cite{litt3} formulated a
secrecy-rate maximization problem of a non-orthogonal unicast-multicast
network, where alternate optimization-based solutions were adopted for
optimizing the non-orthogonal power splitting and RIS's phase shifts. M.\ H.
Khosafa \textit{et al. }inspected the secrecy performance of a RIS-assisted
device-to-device WCS subject to interference from another cellular user \cite%
{lit7}. Also, a comprehensive secrecy analysis for similar RIS-assisted WCSs
in distinct setups was carried out in other work such as \cite%
{lit4,lit5,ris2} and references therein.

\subsection{Motivation}

Although the aforementioned work brought interesting contributions on the
PLS\ analysis and optimization of RIS-aided WCSs, they were constrained by
the assumption of error-free phase shift estimation and quantization. In
fact, reaching a high-precision configuration for the RIS is impractical
\cite{errphase}. The authors in \cite{errphase} provided a bit error
probability analysis of a RIS-assisted system subject to phase estimation
and quantization errors. Importantly, S\'{a}nchez \textit{et al.} assessed
the secrecy performance analysis of a RIS-based network subject to the
presence of phase estimation and quantization errors \cite{vegajavier}.
Furthermore, the work in \cite{lit8,lit10} extended the secrecy analysis of
such a WCS by considering phase estimation and quantization errors along
with multiple eavesdroppers under the presence and absence of a direct link,
respectively. On the other hand, a limited number of work in the literature
analyzed the interplay between conventional dual/multi-hop relaying
techniques and RIS when incorporated together on a transmission system.
Differently from the RIS\ principle, relaying techniques actively process
the signal by either amplifying it (amplify-and-forward\ (AF)), or decoding
it (decode-and-forward (DF)), before handing it to the next relay or
destination node. To this end, work such as \cite{litt4,litt5,litt6} aimed
to amalgamate the benefits of both RIS and active relaying schemes on the
PLS. In \cite{litt4}, the authors assessed the PLS analysis of a three-hop
mixed visible light communication/radio-frequency (RF) WCS whereby a RIS\ is
used to assist the second hop (i.e., RF channel). Furthermore, the authors
of \cite{litt5} aimed at optimizing both RIS\ phase shifts and relay
selection in a multi-relay dual-hop network through deep reinforcement
learning, where a RIS\ was involved to assist both hops. The secrecy level
of a mixed RF-underwater optical WCS\ was quantified in \cite{litt6} with a
single RIS\ assisting the first RF\ hop, under the presence of a single
eavesdropper.

Combining RIS and cooperative relaying schemes has been appealing in a
multi-hop scenario; e.g., wireless mesh network \cite{mesh}; where the
different relaying transceivers can be located in a non-line-of-sight
environment. This motivates the implementation of numerous RISs\ to
accommodate reliable transmissions. On the other hand, the inclusion of
cooperative jamming/artificial noise has been shown to provide significant
improvement in system secrecy. Importantly, several works such as \cite%
{jammingris,lit7} analyzed and discussed the interplay between RIS\ and
jamming schemes in enhancing the PLS of WCSs. Therefore, it is crucial to
provide a comprehensive PLS\ analysis of dual-hop relay-based networks
assisted by RIS along with jamming, by considering phase estimation and/or
quantization errors.

\subsection{Contributions}

Motivated by the above, we aim in this paper at analyzing the secrecy
performance of a dual-hop RIS and jamming-aided WCS. Particularly, two
multi-element RISs are involved to assist the transmission of each of the
two hops, i.e. source-relay and relay-destination, where the DF\ relaying
scheme is implemented at the relay. In addition to this, a malicious
eavesdropper is attempting to overhear the signal broadcasted by the relay
and reflected by the second RIS. Additionally, a jammer is incorporated to
deceive the eavesdropper by injecting a jamming signal, assumed to be
canceled perfectly at the legitimate destination. Lastly, the RIS-assisted
transmission is assumed to be subject to phase quantization errors\ (PQEs).
The current work differs from \cite{vegajavier,lit8,lit10,litt4,litt5,litt6}
where the PLS\ analysis was carried out by considering either perfect phase
estimation and quantization or dual-hop relaying, while the involvement of
cooperative jamming was not considered. To the best of our knowledge, the
current work is the first of its kind to inspect the joint influence of the
DF\ relaying scheme and cooperative jamming along with RIS affected by PQEs
on the PLS. In detail, the main contributions of this work can be summarized
as follows:

\begin{itemize}
\item By virtue of the well-adopted Gamma and Exponential distributions as
accurate approximations, we provide an approximate expression for the
system's intercept probability (IP)\ metric in terms of key setup
parameters, such as the per RIS number of REs, jamming-to-noise power ratio,
legitimate and eavesdropper's links' average signal-to-noise ratios (SNRs),
and the number of quantization bits.

\item We derive an asymptotic expression for the IP\ in the high SNR regime,
whereby the underlying coding gain and diversity order are quantified. The
results show that the secrecy diversity order is proportional to the minimum
of the two RISs' number of REs.

\item We show analytically that the system's secrecy level improves by
increasing the number of REs and/or number of quantization bits. In addition
to this, we prove that the SNRs of the legitimate and wiretap links are
uncorrelated although they share some common terms.

\item We provide extensive numerical and simulation results in order to
quantify the secrecy level of the system versus the various parameters
involved. We found that an IP\ around $10^{-4}$ can be reached with $40$ REs
per RIS and $10$ dB of jamming power-to-noise ratio even when the legitimate
links' average SNRs are $10$ dB below the eavesdropper's one.
\end{itemize}

\subsection{Organization}

The rest of this paper is structured as follows: In Section \ref{sysmodd},
the adopted system model is detailed, while in Section \ref{stats}, we
provide some statistics of the different channels' SNRs.\ In Section \ref%
{expressions}, we derive approximate and high-SNR\ asymptotic expressions
for the IP\, and provide several analytical insights and discussions on the
influence of some key system parameters on the system's security. Numerical
and simulation results are presented and discussed in Section \ref{results}.
Finally, Section \ref{concl}\ concludes the paper.

\subsection{Notations}

Table \ref{nott} lists the symbols and notations used in this paper.
\begin{table*}[t]
\centering\hspace*{-0.2cm}%
\begin{tabular}{c||c||c||c}
\hline\hline
\textbf{\textit{Symbol}} & \textbf{\textit{Meaning}} & \textbf{\textit{Symbol%
}} & \textbf{\textit{Meaning}} \\ \hline\hline
$S$ & Source & $\eta _{i}$ & $L^{(2)}$'s $i$th element phase shift \\ \hline
$R$ & Relay & $\xi _{i}^{(l)}$ & $L^{(l)}$'s $i$th element PQE \\ \hline
$D$ & Destination & $F_{.}^{(c)}\left( \cdot \right) $ & Complementary
cumulative distribution function \\ \hline
$J$ & Jammer & $\Gamma _{\text{inc}}\left( \cdot ,\cdot \right) $ &
Upper-incomplete Gamma function \\ \hline
$E$ & Eavesdropper & $\Gamma \left( \cdot \right) $ & Gamma function \\
\hline
$L^{(l)}$ & $l$-th RIS ($l=1,2$) & $m_{UV}$ & Gamma distribution's shape
parameter for the $U$-$V$ link \\ \hline
$M$ & $L^{(1)}$'s number of REs & $\Omega _{UV}$ & Average equivalent fading
power for the $U$-$V$ link \\ \hline
$N$ & $L^{(2)}$'s number of REs & $\left( \varphi _{l,k}\right) _{l=1,2}$ & $%
k$th moment of the PQEs of $L^{(l)}$ \\ \hline
$\gamma _{UV}$ & $U$-$V$ link's path-loss-normalized instantaneous SNR/SINR
& $\left( n_{b_{l}}\right) _{l}$ & $L^{(l)}$'s number of quantization bits
\\ \hline
$\overline{\gamma }_{UV}$ & $U$-$V$ link's path-loss-normalized average SNR
& $\mathbb{E}\left[ .\right] $ & Expected value \\ \hline
$P_{U}$ & Transmit power of node $U$ & $f_{.}\left( \cdot \right) $ &
Probability density function \\ \hline
$\sigma _{V}^{2}$ & Additive white Gaussian noise power at $V$ & $%
F_{.}\left( \cdot \right) $ & Cumulative distribution function \\ \hline
$h_{XY}$ & $XY$ link's complex-valued fading coefficient & $C_{s}$ & Secrecy
capacity \\ \hline
$\mathcal{CN}\left( \mu ,\sigma ^{2}\right) $ & Complex Gaussian
distribution of mean $\mu $ and variance $\sigma ^{2}$ & $P_{int}$ &
Intercept probability \\ \hline
$\phi _{i}$ & $L^{(1)}$'s $i$th element phase shift & $H_{\cdot ,\cdot
;\cdot ,\cdot ;\cdot ,\cdot }^{\cdot ,\cdot ;\cdot ,\cdot ;\cdot ,\cdot
}\left( \cdot ,\cdot \left\vert \cdot \right. \right) $ & Bivariate Fox's $H$%
-function \\ \hline
\end{tabular}%
\caption{List of symbols.}
\label{nott}
\end{table*}

\section{System Model}

\label{sysmodd}

\begin{figure}[h]
\begin{center}
\includegraphics[scale=0.22]{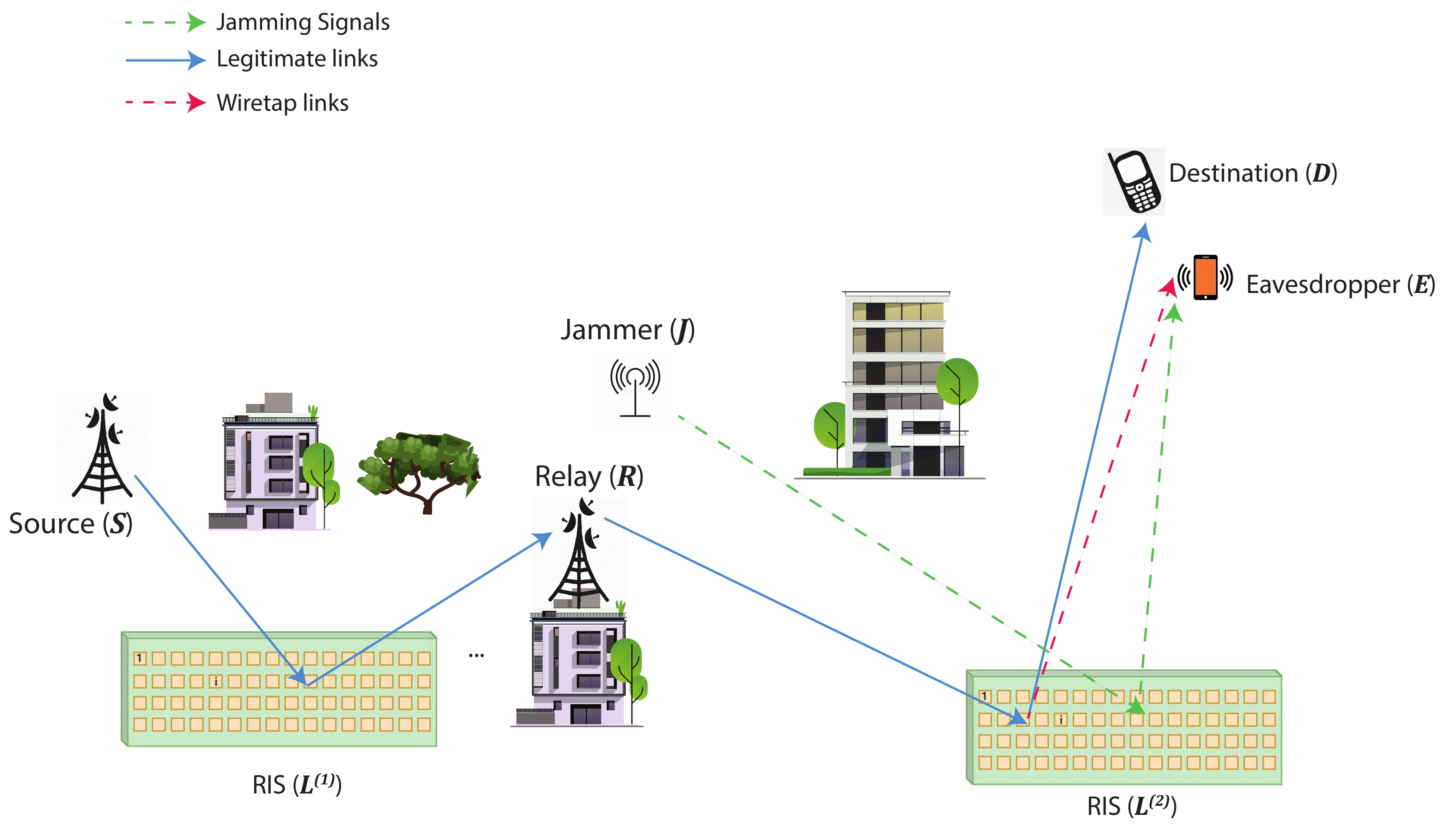}
\end{center}
\caption{System model.}
\label{sysmod}
\end{figure}

We consider a dual-hop\ RIS-assisted wireless network as depicted in Fig. %
\ref{sysmod}, where a source node $\left( S\right) $ communicates with a
destination node $\left( D\right) $ with the help of a DF-based relay $%
\left( R\right)$. Additionally, two RISs, $L^{(1)}$ and $L^{(2)}$, with $M$
and $N$ REs, respectively, are installed to assist the $S$-$R$ and $R$-$D$
transmissions. Furthermore, an eavesdropper ($E$) attempts to compromise the
legitimate link by overhearing the signal broadcasted by $R$. Moreover, a
jammer ($J$) enhances the communication's security level by broadcasting a
jamming signal to disrupt $E$; assuming that $D$ is able to remove such a
jamming signal through some artificial noise cancellation mechanism \cite%
{jamming}. Due to the long distance between the nodes, the direct-link
signals of the $S$-$R$, $R$-$D$, $R$-$E$, $S$-$E$, $J$-$E$, and $S$-$D$
channels can be neglected. Lastly, it is assumed that all the nodes in the
network are equipped with a single antenna.

The received per-hop SNRs at $D$ and $R$ are given as
\begin{equation}
\gamma _{SR}=\overline{\gamma }_{SR}\left\vert
\sum\limits_{i=1}^{M}h_{SL_{i}^{(1)}}h_{L_{i}^{(1)}R}\exp \left( j\phi
_{i}\right) \right\vert ^{2},  \label{snrsri}
\end{equation}%
\begin{equation}
\gamma _{RD}=\overline{\gamma }_{RD}\left\vert
\sum\limits_{i=1}^{N}h_{RL_{i}^{(2)}}h_{L_{i}^{(2)}D}\exp \left( j\eta
_{i}\right) \right\vert ^{2},  \label{snrrdi}
\end{equation}%
respectively, where $\overline{\gamma }_{UV}=\frac{P_{U}}{\sigma _{V}^{2}}$
is the path-loss-normalized average SNR of the $U$-$V$ link with $UV\in
\left\{ SR,RD\right\} $, $P_{U}$ is the transmit power of node $U$, $\sigma
_{R}^{2}$ and $\sigma _{D}^{2}$ are the respective additive white Gaussian
noise powers at $R$ and $D$, respectively, $h_{SL_{i}^{(1)}}$, $%
h_{L_{i}^{(1)}R}$, $h_{RL_{i}^{(2)}}$, and $h_{L_{i}^{(2)}D}$ are the\
channel fading coefficients of the $S$-$L_{i}^{(1)}$, $L_{i}^{(1)}$-$R$, $R$-%
$L_{i}^{(2)}$ and $L_{i}^{(2)}$-$D$ links, respectively, assumed to be
independent and identically distributed zero-mean complex Gaussian random
variables (RVs) with distribution $\mathcal{CN}\left( 0,1\right) $, i.e.,
Rayleigh fading, where $L_{i}^{(l)}$ $(l=1,2)$\ is $L^{(l)}$'s $i$th RE, $j=%
\sqrt{-1}$, and $\phi _{i}$ and $\eta _{i}$ are the phase shifts of $L^{(1)}$
and $L^{(2)}$'s $i$th RE, respectively. On the other hand, the instantaneous
signal-to-interference-and-noise ratio (SINR) at $E$ can be written as%
\begin{equation}
\gamma _{RE}=\frac{\overline{\gamma }_{RE}\left\vert
\sum\limits_{i=1}^{N}h_{RL_{i}^{(2)}}h_{L_{i}^{(2)}E}\exp \left( j\eta
_{i}\right) \right\vert ^{2}}{\overline{\gamma }_{JE}\left\vert
\sum\limits_{i=1}^{N}h_{JL_{i}^{(2)}}h_{L_{i}^{(2)}E}\exp \left( j\eta
_{i}\right) \right\vert ^{2}+1},  \label{snrrei}
\end{equation}%
where $h_{JL_{i}^{(2)}}$ is the fading coefficient of the $J$-$L_{i}^{(2)}$
link.

Furthermore, we assume a perfect channel state information (CSI)\ estimation
at the two RISs\footnote{%
The end-to-end legitimate channels estimation ($S$-$L^{(1)}$-$R$ and $R$-$%
L^{(2)}$-$D$) is performed at $R$ and $D$, respectively, by sending pilot
symbols from $S$ and $R$ over various RIS\ phase configurations, whereby the
cascaded channel can be estimated linearly by combining the corresponding
received signals. Afterward, the channel coefficients are fed back to the
RISs' controllers, by which appropriate phase shifts can be selected
accordingly \cite{rischannel}.}, while the estimated phase is subject to
quantization errors. To this end, the phase shifts of the $i$-th element of $%
L^{(l)}$($l=1,2$) are given by \cite{errphase}
\begin{equation}
\phi _{i}=-\left[ \arg \left( h_{SL_{i}^{(1)}}\right) +\arg \left(
h_{L_{i}^{(1)}R}\right) \right] +\xi _{i}^{(1)},1\leq i\leq M  \label{phii}
\end{equation}%
and
\begin{equation}
\eta _{i}=-\left[ \arg \left( h_{RL_{i}^{(2)}}\right) +\arg \left(
h_{L_{i}^{(2)}D}\right) \right] +\xi _{i}^{(2)},1\leq i\leq N,  \label{etai}
\end{equation}%
respectively, where $\xi _{i}^{(l)}$ are the corresponding PQE at the $i$-th
elements of $L^{(l)}$ $\left( l=1,2\right) $. As a result, the corresponding
per-hop SNRs at $R\ $and $D$ and the SINR at $E$, given by (\ref{snrsri}), (%
\ref{snrrdi}), and (\ref{snrrei}), respectively, become%
\begin{equation}
\gamma _{SR}=\overline{\gamma }_{SR}\left\vert
\sum\limits_{i=1}^{M}\left\vert h_{SL_{i}^{(1)}}\right\vert \left\vert
h_{L_{i}^{(1)}R}\right\vert \exp \left( j\xi _{i}^{(1)}\right) \right\vert
^{2},  \label{snrsr}
\end{equation}%
\begin{equation}
\gamma _{RD}=\overline{\gamma }_{RD}\left\vert
\sum\limits_{i=1}^{N}\left\vert h_{RL_{i}^{(2)}}\right\vert \left\vert
h_{L_{i}^{(2)}D}\right\vert \exp \left( j\xi _{i}^{(2)}\right) \right\vert
^{2},  \label{snrrd}
\end{equation}%
and%
\begin{equation}
\gamma _{RE}=\frac{\underset{\triangleq \Gamma _{RE}}{\underbrace{\overline{%
\gamma }_{RE}\left\vert \sum\limits_{i=1}^{N}\left\vert
h_{RL_{i}^{(2)}}\right\vert \left\vert h_{L_{i}^{(2)}E}\right\vert \exp
\left( j\varpi _{i}\right) \right\vert ^{2}}}}{\underset{\triangleq \Gamma
_{JE}}{\underbrace{\overline{\gamma }_{JE}\left\vert
\sum\limits_{i=1}^{N}h_{JL_{i}^{(2)}}h_{L_{i}^{(2)}E}\exp \left( j\eta
_{i}\right) \right\vert ^{2}}}+1},  \label{snrre}
\end{equation}%
with
\begin{equation}
\varpi _{i}=\arg \left( h_{L_{i}^{(2)}E}\right) -\arg \left(
h_{L_{i}^{(2)}D}\right) +\xi _{i}^{(2)}.  \label{omegai}
\end{equation}

\section{Statistical Properties}

\label{stats}

The legitimate instantaneous SNR\ expressions in (\ref{snrsr}) and (\ref%
{snrrd}) are incorporating the residual PQEs of $L^{(1)}$ and $L^{(2)}$,
respectively. For a given number of quantization bits, such SNRs\ can be
accurately approximated by a Gamma distribution where the approximate
complementary cumulative distribution function (CCDF) can be derived by
virtue of integrating the probability density function (PDF) in \cite[Eq.
(13)]{errphase} as follows%
\begin{equation}
F_{\gamma _{UV}}^{(c)}\left( z\right) \approx \frac{\Gamma _{\text{inc}%
}\left( m_{UV},\frac{m_{UV}}{\Omega _{UV}\overline{\gamma }_{UV}}z\right) }{%
\Gamma \left( m_{UV}\right) },UV\in \{SR,RD\},  \label{ccdf}
\end{equation}%
with $\Gamma _{\text{inc}}\left( \cdot ,\cdot \right) $ and $\Gamma \left(
\cdot \right) $ are the upper-incomplete and complete Gamma functions,
respectively \cite[Eqs (8.350.2, 8.310.1)]{integrals}. Furthermore, by
considering the Rayleigh fading model \footnote{%
The channel between the various nodes and the RIS elements can be
represented by the Rayleigh fading model, as was considered in various
studies such as \cite{lit1,lit2,electlett,basar}. Such a model can be used
in the considered case of the absence of a line-of-sight link between the
nodes and the RIS.} over all channels, we have $m_{UV}=\frac{\mathcal{K}}{2}%
\frac{\varphi _{l,1}^{2}\frac{\pi ^{2}}{16}}{1+\varphi _{l,2}-2\varphi
_{l,1}^{2}\frac{\pi ^{2}}{16}}$, $\mathcal{K=}\left\{
\begin{array}{c}
M,\text{ if }UV=SR \\
N\text{, if }UV=RD%
\end{array}%
\right. $, $l\mathcal{=}\left\{
\begin{array}{c}
1,\text{ if }UV=SR \\
2\text{, if }UV=RD%
\end{array}%
\right. $, $\varphi _{l,k}=\mathbb{E}\left[ \exp \left( jk\xi
_{i}^{(l)}\right) \right] $, and $\Omega _{UV}=\left( \frac{\mathcal{K}\pi
\varphi _{l,1}}{4}\right) ^{2}$ \cite{errphase}. When only a finite set of $%
2^{n_{b_{l}}}$ phases can be configured at the $l$th RIS, with $n_{b_{l}}$
being the corresponding number of quantization bits, the PQE $\xi _{i}^{(l)}$
is uniformly distributed over the interval $\left[ -\frac{\pi }{2^{n_{b_{l}}}%
},\frac{\pi }{2^{n_{b_{l}}}}\right] $ with $\varphi _{l,k}=\frac{%
2^{n_{b_{l}}+1-k}\sin \left( 2^{k-1-n_{b_{l}}}\pi \right) }{\pi }$ \cite%
{errphase}.

\begin{remark}
The per-hop SNRs ($\gamma _{SR}$ and $\gamma _{RD}$) are approximated by
Gamma-distributed RVs with shape parameter%
\begin{equation}
m_{UV}=\frac{\mathcal{K}}{2}\frac{\varphi _{l,1}^{2}\frac{\pi ^{2}}{16}}{%
1+\varphi _{l,2}-2\varphi _{l,1}^{2}\frac{\pi ^{2}}{16}},UV\in \left\{
SR,RD\right\} ,  \label{shape}
\end{equation}%
and scale parameter $\frac{\Omega _{UV}\overline{\gamma }_{UV}}{m_{UV}}$ with%
\begin{equation}
\Omega _{UV}=\frac{\mathcal{K}^{2}}{16}\left( \frac{\pi \sin \left( X\right)
}{X}\right) ^{2},X\in \left[ 0,\frac{\pi }{2}\right] ,  \label{omega}
\end{equation}%
where $X=\frac{\pi }{2^{n_{b_{l}}}}$ and $n_{b_{l}}\in \left[ 1,\infty %
\right[ $. On the other hand, the average value of $\gamma _{UV}$ is the
product of its shape and scale parameters, i.e.,
\begin{equation}
\mathbb{E}\left[ \gamma _{UV}\right] =\Omega _{UV}\overline{\gamma }_{UV},
\label{gammabar}
\end{equation}%
which is proportional to $\Omega _{UV}$.

\begin{enumerate}
\item By differentiating $\Omega _{UV}$ with respect to $X$, one obtains%
\begin{equation}
\frac{\partial \Omega _{UV}}{\partial X}=\frac{\mathcal{K}^{2}\pi ^{2}\sin
\left( X\right) \cos \left( X\right) }{8X^{3}}\left( X-\tan \left( X\right)
\right) .
\end{equation}%
As $\sin \left( X\right) \geq 0$ and $\cos \left( X\right) \geq 0$ for $X\in %
\left[ 0,\frac{\pi }{2}\right] $, the above derivative's sign depends on the
one of $g(X)=X-\tan \left( X\right) $. To this end, we have the following%
\begin{equation}
g^{\prime }(X)=1-\sec ^{2}\left( X\right) ,
\end{equation}%
where $\sec \left( \cdot \right) =\frac{1}{\cos \left( \cdot \right) }$ is
the secant function, which is obviously greater than or equal to $1$ for $%
X\in \left[ 0,\frac{\pi }{2}\right] $. Hence, $g^{\prime }(X)\leq 0$ over
this latter interval which yields that (i) $g(X)$ is a decreasing function
over the same interval. Furthermore, we have (ii)
\begin{eqnarray}
\lim_{x\rightarrow 0}g(X) &=&0, \\
\lim_{x\rightarrow \frac{\pi }{2}}g(X) &=&-\infty .
\end{eqnarray}%
Thus, leveraging (i)\ and (ii), $g(X)\leq 0$ for $X\in \left[ 0,\frac{\pi }{2%
}\right] $. Therefore, $\Omega _{UV}$ is a decreasing function of $X$. On
the other hand, it is obvious that $X$ is inversely proportional to $%
n_{b_{l}}$. As a result, $\Omega _{UV}$ is increasing in terms of $n_{b_{l}}$%
. Thus, the higher the number of quantization bits, the greater $\Omega
_{UV} $, leading to higher average value of the per-hop legitimate SNR,
i.e., $\mathbb{E}\left[ \gamma _{UV}\right] $.

\item We can infer from (\ref{omega}) and (\ref{gammabar}) that the average
legitimate SNRs, i.e., $\mathbb{E}\left[ \gamma _{SR}\right] $ and $\mathbb{E%
}\left[ \gamma _{RD}\right] $, scale with $M^{2}$ and $N^{2}$, respectively.
Therefore, the higher the number of REs, the greater the legitimate per hop
SNRs.
\end{enumerate}
\end{remark}

Since $R$ performs DF\ relaying protocol, the end-to-end (e2e) SNR can be
written as \cite[Eq. (13)]{sustainable}%
\begin{equation}
\gamma _{eq}=\min \left( \gamma _{SR},\gamma _{RD}\right) ,  \label{gammaeq}
\end{equation}%
where the underlying CCDF\ is given as \
\begin{equation}
F_{\gamma _{eq}}^{(c)}\left( x\right) =F_{\gamma _{SR}}^{(c)}\left( x\right)
F_{\gamma _{RD}}^{(c)}\left( x\right) .  \label{ccdfeq}
\end{equation}

It has been demonstrated in \cite{errphase,vegajavier} that the random
variables (RVs) $\Gamma _{RE}$ and $\Gamma _{JE}$, given in the numerator
and denominator of (\ref{snrre}), respectively, can be approximated by an
Exponential distribution with PDF and cumulative distribution function (CDF)
expressed as%
\begin{equation}
f_{\Gamma _{UE}}\left( z\right) \approx \frac{1}{N\overline{\gamma }_{UE}}%
\exp \left( -\frac{z}{N\overline{\gamma }_{UE}}\right) ,U\in \left\{
R,J\right\}  \label{pdfpart}
\end{equation}%
and%
\begin{equation}
F_{\Gamma _{UE}}\left( z\right) \approx 1-\exp \left( -\frac{z}{N\overline{%
\gamma }_{UE}}\right) ,U\in \left\{ R,J\right\} ,  \label{cdfpart}
\end{equation}%
respectively, with an average value of%
\begin{equation}
\mathbb{E}\left[ \Gamma _{UE}\right] =N\overline{\gamma }_{UE},U\in \left\{
R,J\right\} .  \label{gammauebar}
\end{equation}%
As a consequence, the CDF\ of $\gamma _{RE}$, defined by (\ref{snrre}), can
be computed as follows%
\begin{align}
F_{\gamma _{RE}}\left( z\right) & \triangleq \int_{0}^{\infty }F_{\Gamma
_{RE}}\left( z\left( x+1\right) \right) f_{\Gamma _{JE}}\left( x\right) dx
\label{stp1} \\
& \overset{(a)}{\approx }\frac{1}{N\overline{\gamma }_{JE}}\int_{0}^{\infty
}\left( 1-\exp \left( -\frac{z\left( x+1\right) }{N\overline{\gamma }_{RE}}%
\right) \right)  \notag \\
& \times \exp \left( -\frac{x}{N\overline{\gamma }_{JE}}\right) dx, \\
& \overset{(b)}{=}1-\frac{\exp \left( -\frac{z}{N\overline{\gamma }_{SE}}%
\right) }{\frac{\overline{\gamma }_{JE}}{\overline{\gamma }_{SE}}z+1}.
\label{stpb}
\end{align}

By incorporating (\ref{pdfpart}) with $U=J$ and (\ref{cdfpart}) with $U=R$
into (\ref{stp1}), \textit{Step} \textit{(a)} is reached, while \textit{Step
(b)}\ is formed relying on the change of variable $t=x+1$ and the integral
of the exponential function.

Consequently, the respective PDF\ is readily obtained by a differentiation
of (\ref{stpb}) with respect to $z$ as
\begin{equation}
f_{\gamma _{RE}}\left( z\right) \approx \frac{\exp \left( -\frac{z}{N%
\overline{\mathcal{\gamma }}_{RE}}\right) \left[ \overline{\mathcal{\gamma }}%
_{JE}N\overline{\mathcal{\gamma }}_{RE}+\overline{\mathcal{\gamma }}_{RE}+%
\overline{\mathcal{\gamma }}_{JE}z\right] }{N\left( \overline{\mathcal{%
\gamma }}_{RE}+\overline{\mathcal{\gamma }}_{JE}z\right) ^{2}}.  \label{pdf}
\end{equation}

\begin{remark}
Leveraging (\ref{gammauebar}), the average value of the eavesdropper link's
SINR (i.e., $\gamma _{RE}$), given by (\ref{snrre}), can be written as
\begin{eqnarray}
\mathbb{E}\left[ \gamma _{RE}\right] &=&\frac{\mathbb{E}\left[ \Gamma _{RE}%
\right] }{\mathbb{E}\left[ \Gamma _{JE}\right] +1}  \notag \\
&=&\frac{N\overline{\mathcal{\gamma }}_{RE}}{N\overline{\mathcal{\gamma }}%
_{JE}+1}.
\end{eqnarray}

Importantly, in the absence of jamming (i.e., $\overline{\mathcal{\gamma }}%
_{JE}=0$), which is among optimal scenarios for the eavesdropper, one
obtains $\mathbb{E}\left[ \gamma _{RE}\right] =N\overline{\mathcal{\gamma }}%
_{RE}$, which scales with $N$. On the other hand, when $\overline{\mathcal{%
\gamma }}_{RE}=\overline{\mathcal{\gamma }}_{JE}\rightarrow \infty ,$ one
obtains $\mathbb{E}\left[ \gamma _{RE}\right] =1,$ which scales
independently of $N$.
\end{remark}

\begin{lemma}[Independence of the Relay-Destination, Jammer-Eavesdropper,
and Relay-Eavesdropper's Cascaded Fading]
The cascaded fading coefficients of the legitimate second hop, eavesdropper,
and jammer-eavesdropper links, given by
\begin{equation}
g_{RD}=\sum\limits_{i=1}^{N}\left\vert h_{RL_{i}^{(2)}}\right\vert
\left\vert h_{L_{i}^{(2)}D}\right\vert \exp \left( j\xi _{i}^{(2)}\right) ,
\label{grd}
\end{equation}%
\begin{equation}
g_{RE}=\sum\limits_{i=1}^{N}\left\vert h_{RL_{i}^{(2)}}\right\vert
\left\vert h_{L_{i}^{(2)}E}\right\vert \exp \left( j\varpi _{i}\right) ,
\label{gre}
\end{equation}%
and\newline
\begin{equation}
g_{JE}=\sum\limits_{i=1}^{N}h_{JL_{i}^{(2)}}h_{L_{i}^{(2)}E}\exp \left(
j\eta _{i}\right) ,  \label{gje}
\end{equation}%
\qquad respectively, are mutually independent.

\begin{IEEEproof}
The proof is provided in Appendix A.
\end{IEEEproof}
\end{lemma}

\section{PHY\ Security Analysis}

\label{expressions}

In this section, novel approximate and high-SNR-asymptotic expressions for
the secrecy IP are derived.

The PLS\ analysis is based on the secrecy capacity metric, which is the
maximal achievable transmission rate satisfying a certain equivocation rate
at the undesired receiver \cite{barros}. Mathematically, it is defined as
the difference between the legitimate and illegitimate links' capacities as
follows%
\begin{equation}
C_{s}=C_{l}-C_{e},  \label{Cs}
\end{equation}%
where%
\begin{equation}
C_{l}=\log _{2}\left( 1+\min \left( \gamma _{SR},\gamma _{RD}\right) \right)
,  \label{Cl}
\end{equation}%
and%
\begin{equation}
C_{e}=\log _{2}\left( 1+\gamma _{RE}\right) ,  \label{Ce}
\end{equation}%
are the legitimate and eavesdropper's channels capacities, respectively.

To this end, the IP metric represents the probability of the event when the
eavesdropper's channel capacity is above the legitimate's, which can be
written as \cite[Eq. (24)]{ojcoms}%
\begin{align}
P_{int}& \triangleq \Pr \left[ C_{s}<0\right]  \notag \\
& =1-\Pr \left[ \min \left( \gamma _{SR},\gamma _{RD}\right) \geq \gamma
_{RE}\right] .  \label{ipdef}
\end{align}

\begin{remark}
The overall secrecy capacity in (\ref{Cs}) can be expressed as
\begin{equation}
C_{s}=\min \left( C_{SR},C_{RD}\right) ,  \label{Csp}
\end{equation}%
where%
\begin{equation}
C_{UV}=\log _{2}\left( \frac{1+\gamma _{UV}}{1+\gamma _{RE}}\right) ,UV\in
\left\{ SR,RD\right\} .  \label{Cshop}
\end{equation}%
\newline

\begin{enumerate}
\item As pointed out in \textit{Remark 1.1, }the average legitimate SNR\ per
hop, i.e., $\mathbb{E}\left[ \gamma _{UV}\right] $, is an increasing
function of $n_{b_{l}}$, while the eavesdropper's average SINR ($\mathbb{E}%
\left[ \gamma _{RE}\right] $) is independent of it, as shown in \textit{Remark 2}. Thus, the higher $%
n_{b_{l}}$, the greater the SNR\ per-hop, leading to an improved per-hop
(eq. (\ref{Cshop})) and e2e (eq. (\ref{Csp})) secrecy capacities. As a
consequence, the system's IP, given by (eq. (\ref{ipdef})), decreases, i.e.,
better system secrecy.

\item Furthermore, $\mathbb{E}\left[ \gamma _{UV}\right] $ scales with $%
M^{2} $ and $N^{2}$ for the $S$-$R$ and $R$-$D$ hops, respectively, as
discussed in \textit{Remark 1.2}, while $\mathbb{E}\left[ \gamma _{RE}\right]
$ scales with $N$ in an inadequate scenario for the legitimate nodes (i.e.,
absence of jamming). Therefore, the ratio of $\mathbb{E}\left[ \gamma _{UV}%
\right] $ and $\mathbb{E}\left[ \gamma _{RE}\right] $ is expressed as%
\begin{equation}
\frac{\mathbb{E}\left[ \gamma _{UV}\right] }{\mathbb{E}\left[ \gamma _{RE}%
\right] }=\frac{\mathcal{K}^{2}\overline{\gamma }_{UV}2^{2n_{b_{l}}}\sin
^{2}\left( \frac{\pi }{2^{n_{b_{l}}}}\right) }{16N\overline{\gamma }_{RE}},
\end{equation}%
with $\mathcal{K=}\left\{
\begin{array}{c}
M,\text{ if }UV=SR \\
N\text{, if }UV=RD%
\end{array}%
\right. $. From another front, we have the following function%
\begin{equation}
h\left( y\right) =y^{2}\sin ^{2}\left( \frac{\pi }{y}\right) ,
\end{equation}%
where $y=2^{n_{b_{l}}}\in \left[ 2,\infty \right[ $, and the derivative of $%
h\left( y\right) $ is%
\begin{equation}
h^{\prime }\left( y\right) =2y\sin \left( \frac{\pi }{y}\right) \left[ \sin
\left( \frac{\pi }{y}\right) -\frac{\pi }{y}\cos \left( \frac{\pi }{y}%
\right) \right] .
\end{equation}%
\qquad\ Obviously, the sign of $h^{\prime }(y)$ is the one of $\sin \left(
\frac{\pi }{y}\right) -\frac{\pi }{y}\cos \left( \frac{\pi }{y}\right) $.
Let us solve the inequality%
\begin{equation}
\sin \left( \frac{\pi }{y}\right) -\frac{\pi }{y}\cos \left( \frac{\pi }{y}%
\right) >0\Leftrightarrow \tan \left( \frac{\pi }{y}\right) >\frac{\pi }{y},
\end{equation}%
which holds for $y\in \left[ 2,\infty \right[ $ as has been shown in \textit{%
Remark 1.1}. Thus, $h^{\prime }\left( y\right) \geq 0$ which yields that $%
h\left( y\right) $ increases with $y$, and consequently with $n_{b_{l}}$.
Henceforth, this yields that $\frac{2^{2n_{b_{l}}}\sin ^{2}\left( \frac{\pi
}{2^{n_{b_{l}}}}\right) }{16}\geq \frac{1}{4}$ for $n_{b_{l}}\geq 1$. To
this end, when $\overline{\gamma }_{UV}=\overline{\gamma }_{RE}$, we have
the following%
\begin{equation}
\frac{\mathbb{E}\left[ \gamma _{UV}\right] }{\mathbb{E}\left[ \gamma _{RE}%
\right] }\geq \frac{\mathcal{K}^{2}}{4N}.  \label{ratio}
\end{equation}%
Henceforth, (\ref{ratio}) shows that raising the number of REs for $\mathcal{%
K>}4$ can boost the legitimate-to-wiretap average SNRs ratio. Consequently,
the per-hop and e2e secrecy capacities, given by (\ref{Cshop}) and (\ref{Csp}%
), respectively, increase. As a result, the IP\ decreases, i.e., enhanced
security.

\item It can be obviously noted from (\ref{ccdf}) and (\ref{ccdfeq}) that
for $M=N$ and $n_{b_{1}}=n_{b_{2}}$ $($i.e., $m_{SR}=m_{RD}$, $\Omega
_{SR}=\Omega _{RD})$, the CCDF of the end-to-end SNR\ manifests a symmetric
behavior with respect to $\overline{\gamma }_{SR}$ and $\overline{\gamma }%
_{RD}$, i.e., $\left( F_{\gamma _{eq}}^{(c)}\left( x\right) \right) _{%
\overline{\gamma }_{SR}=\alpha ,\overline{\gamma }_{RD}=\beta }=\left(
F_{\gamma _{eq}}^{(c)}\left( x\right) \right) _{\overline{\gamma }%
_{SR}=\beta ,\overline{\gamma }_{RD}=\alpha }$ , with $\alpha ,\beta >0$.
Furthermore, the IP in (\ref{ipdef}) can be developed as follows%
\begin{eqnarray}
P_{int} &=&1-\Pr \left[ \min \left( \gamma _{SR},\gamma _{RD}\right) \geq
\gamma _{RE}\right]  \notag \\
&=&1-\int_{0}^{\infty }F_{\gamma _{eq}}^{(c)}\left( x\right) f_{\gamma
_{RE}}\left( x\right) dx,
\end{eqnarray}%
which exhibits as well a symmetric behavior with respect to $\overline{%
\gamma }_{SR}$ and $\overline{\gamma }_{RD}$.\qquad \qquad \qquad \qquad
\qquad \qquad
\end{enumerate}
\end{remark}

\subsection{Exact Analysis}

\begin{proposition}
The IP\ of the considered dual-hop RIS-aided WCS\ can be written in terms of
the approximate form given by (\ref{ipfinal}) at the top of the next page,
where $%
H_{p_{1},q_{1};p_{2},p_{2};p_{3},q_{3}}^{m_{1},n_{1};m_{2},n_{2};m_{3},n_{3}}\left( \cdot ,\cdot \left\vert \cdot \right. \right)
$ is the bivariate Fox's $H$-function \cite[Eq. (10.1)]{yakub}, and $\Lambda
_{i}\left( k,l\right) $ $\left( i=1,2\right) $ are given by (\ref{dlt1}) and
(\ref{dlt2}).
\begin{figure*}[t]
{\normalsize 
\setcounter{mytempeqncnt}{\value{equation}}
\setcounter{equation}{42} }%
\begin{equation}
P_{int}\approx 1-\frac{1}{\Gamma \left( m_{SR}\right) \Gamma \left(
m_{RD}\right) }\sum\limits_{l=0}^{\infty }\left( \frac{-1}{N\overline{%
\mathcal{\gamma }}_{JE}}\right) ^{l}\left[
\begin{array}{c}
\sum\limits_{k=1}^{2}\frac{1}{l!\left( N\overline{\mathcal{\gamma }}%
_{JE}\right) ^{2-k}}H_{2,0;1,2;1,2}^{0,2;2,0;2,0}\left( \frac{m_{SR}%
\overline{\gamma }_{RE}}{\Omega _{SR}\overline{\mathcal{\gamma }}_{SR}%
\overline{\gamma }_{JE}},\frac{m_{RD}\overline{\gamma }_{RE}}{\Omega _{RD}%
\overline{\mathcal{\gamma }}_{RD}\overline{\gamma }_{JE}}\left\vert \Lambda
_{1}\left( k,l\right) \right. \right) \\
\frac{1}{N\overline{\mathcal{\gamma }}_{JE}}\sum\limits_{k=1}^{2}\left(
1+l\right) ^{k-1}H_{1,0;1,2;1,2}^{0,1;2,0;2,0}\left( \frac{Nm_{SR}\overline{%
\mathcal{\gamma }}_{RE}}{\Omega _{SR}\overline{\mathcal{\gamma }}_{SR}},%
\frac{Nm_{RD}\overline{\mathcal{\gamma }}_{RE}}{\Omega _{RD}\overline{%
\mathcal{\gamma }}_{RD}}\left\vert \Lambda _{2}\left( k,l\right) \right.
\right)%
\end{array}%
\right] .  \label{ipfinal}
\end{equation}%
\par
{\normalsize 
\hrulefill 
\vspace*{4pt} }
\end{figure*}
\begin{figure*}[t]
{\normalsize 
\setcounter{mytempeqncnt}{\value{equation}}
\setcounter{equation}{43} }%
\begin{align}
\Lambda _{1}\left( k,l\right) & =\left(
\begin{array}{c}
\left( 2-k+l,-1,-1\right) ,\left( -l,1,1\right) ;-:-;\left( 1,1\right)
:-;\left( 1,1\right) \\
-:\left( 0,1\right) ,\left( m_{SR},1\right) ;-:\left( 0,1\right) ,\left(
m_{RD},1\right) ;-%
\end{array}%
\right) .  \label{dlt1} \\
\Lambda _{2}\left( k,l\right) & =\left(
\begin{array}{c}
\left( k+l,1,1\right) ;-:-;\left( 1,1\right) :-;\left( 1,1\right) \\
-:\left( 0,1\right) ,\left( m_{SR},1\right) ;-:\left( 0,1\right) ,\left(
m_{RD},1\right) ;-%
\end{array}%
\right) .  \label{dlt2}
\end{align}%
\par
{\normalsize 
\hrulefill 
\vspace*{4pt} }
\end{figure*}

\begin{IEEEproof}
The proof is provided in Appendix B.
\end{IEEEproof}
\end{proposition}

\subsection{Asymptotic Analysis}

\begin{proposition}
At the high SNR\ regime ($\overline{\mathcal{\gamma }}_{SR}=\overline{%
\mathcal{\gamma }}_{RD}=\overline{\mathcal{\gamma }}\rightarrow \infty $),
the IP\ of the considered dual-hop RIS-assisted WCS\ can be asymptotically
expanded as
\begin{equation}
P_{int}^{\left( \infty \right) }\sim G_{c}\overline{\gamma }^{-G_{d}},
\label{ipass}
\end{equation}%
where $G_{c}$ is defined in (\ref{Gc}) shown at the top of the next page,
while
\begin{figure*}[t]
{\normalsize 
\setcounter{mytempeqncnt}{\value{equation}}
\setcounter{equation}{46} }
\par
\begin{equation}
G_{c}=\left\{
\begin{array}{l}
\frac{\left( \frac{m_{SR}\overline{\mathcal{\gamma }}_{RE}}{\Omega _{SR}%
\overline{\mathcal{\gamma }}_{JE}}\right) ^{m_{SR}}\Gamma \left( 1-m_{SR},%
\frac{1}{N\overline{\gamma }_{JE}}\right) }{\exp \left( -\frac{1}{N\overline{%
\gamma }_{JE}}\right) }\text{; }M<N \\
\frac{\left( \frac{m_{RD}\overline{\mathcal{\gamma }}_{RE}}{\Omega _{RD}%
\overline{\mathcal{\gamma }}_{JE}}\right) ^{m_{RD}}\Gamma \left( 1-m_{RD},%
\frac{1}{N\overline{\gamma }_{JE}}\right) }{\exp \left( -\frac{1}{N\overline{%
\gamma }_{JE}}\right) }\text{; }M>N \\
\frac{\left( \frac{m_{SR}\overline{\mathcal{\gamma }}_{RE}}{\Omega _{SR}%
\overline{\mathcal{\gamma }}_{JE}}\right) ^{m_{SR}}\Gamma \left( 1-m_{SR},%
\frac{1}{N\overline{\gamma }_{JE}}\right) }{\exp \left( -\frac{1}{N\overline{%
\gamma }_{JE}}\right) }+\frac{\left( \frac{m_{RD}\overline{\mathcal{\gamma }}%
_{RE}}{\Omega _{RD}\overline{\mathcal{\gamma }}_{JE}}\right) ^{m_{RD}}\Gamma
\left( 1-m_{RD},\frac{1}{N\overline{\mathcal{\gamma }}_{JE}}\right) }{\exp
\left( -\frac{1}{N\overline{\gamma }_{JE}}\right) }\text{; }M=N%
\end{array}%
\right. .  \label{Gc}
\end{equation}%
\par
{\normalsize 
\hrulefill 
\vspace*{4pt} }
\end{figure*}
\begin{equation}
G_{d}=\min \left( m_{SR},m_{RD}\right) .  \label{Gd}
\end{equation}

\begin{IEEEproof}
The proof is provided in Appendix C.
\end{IEEEproof}
\end{proposition}

\begin{remark}
Leveraging (\ref{shape}), it can be noted that the secrecy diversity order,
given by (\ref{Gd}), is directly proportional to the minimum among the
number of REs of $L^{(1)}$ and $L^{(2)}$.
\end{remark}

\section{Numerical Results}

\label{results}

In this section, we provide numerical results for the secrecy level of the
analyzed dual-hop RIS-assisted WCS. The system parameters' default values
are specified in Table \ref{param}. Furthermore, throughout the simulation
results, the notation $\overline{\gamma }$ is used when the IP\ is plotted
vs $\overline{\gamma }_{SR}$ with $\overline{\gamma }_{RD}=\overline{\gamma }%
_{SR}=\overline{\gamma }$. Furthermore, we generated $3\times 10^{6}$ random
values for the per-hop fading coefficients, i.e. $h_{SL^{(1)}},$ $%
h_{L^{(1)}R},$ $h_{RL^{(2)}},$ $h_{JL^{(2)}},$ $h_{L^{(2)}D}, $ $%
h_{L^{(2)}E} $, in order to perform Monte Carlo simulations. Lastly, the
simulation results shown in the two-dimensions figures (Except Fig. \ref%
{fig6}) are represented by x-shaped markers, while analytical results are
illustrated by solid lines with different geometric shapes.
\begin{table}[tbp]
\centering%
\begin{tabular}{c||c}
\hline
\textbf{\textit{Parameter}} & \textbf{\textit{Value}} \\ \hline
$\overline{\gamma }_{SR},\overline{\gamma }_{RD}$ & $30$ dB \\ \hline
$n_{b_{1}},n_{b_{2}}$ & $3$ bits \\ \hline
$\overline{\gamma }_{RE}$ & $40$ dB \\ \hline
$\overline{\gamma }_{JE}$ & $10$ dB \\ \hline
$M,N$ & $32$%
\end{tabular}%
\caption{Simulation parameters' values.}
\label{param}
\end{table}
\begin{figure}[tbp]
\begin{center}
\includegraphics[scale=.58]{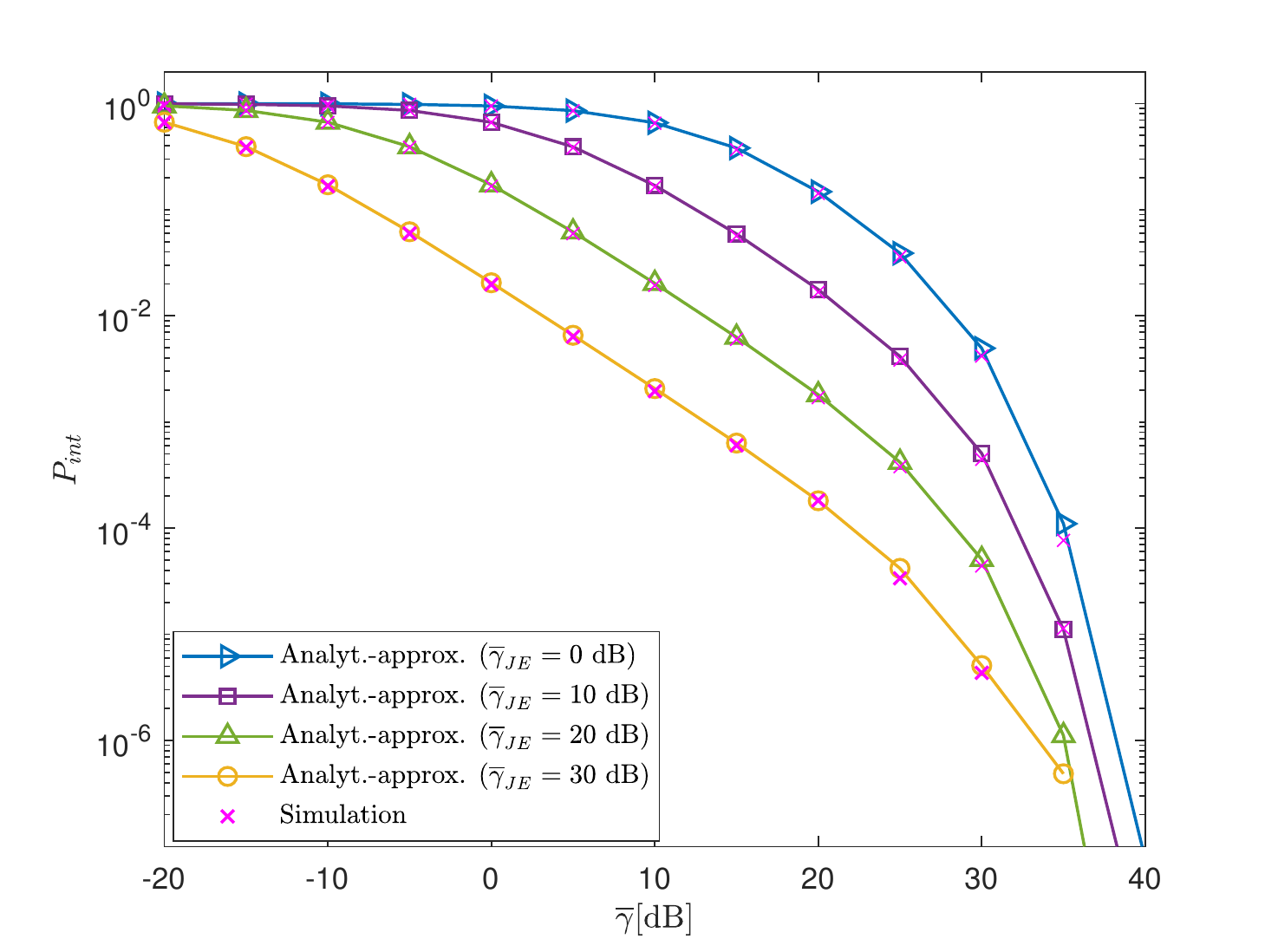}
\end{center}
\caption{IP versus $\overline{\protect\gamma }$ for various $\overline{%
\protect\gamma }_{JE}$ values.}
\label{fig1}
\end{figure}

In Fig. \ref{fig1}, the IP of the considered dual-hop WCS is shown versus $%
\overline{\gamma }$. One can remark that the solid lines, corresponding to
the derived approximate form in (\ref{ipfinal}), tightly match the
simulation results, particularly for $\overline{\gamma }\leq 30$ dB.
Furthermore, the IP exhibits a remarkable decrease vs. $\overline{\gamma }$
as expected. Finally, one can ascertain that the higher the jamming
power-to-noise ratio, the better the secrecy is, where an IP of $10^{-6}$
can be reached with $\overline{\gamma }_{JE}=20$ dB, though the legitimate
average SNRs are below the eavesdropper's one.
\begin{figure}[tbp]
\begin{center}
\includegraphics[scale=.58]{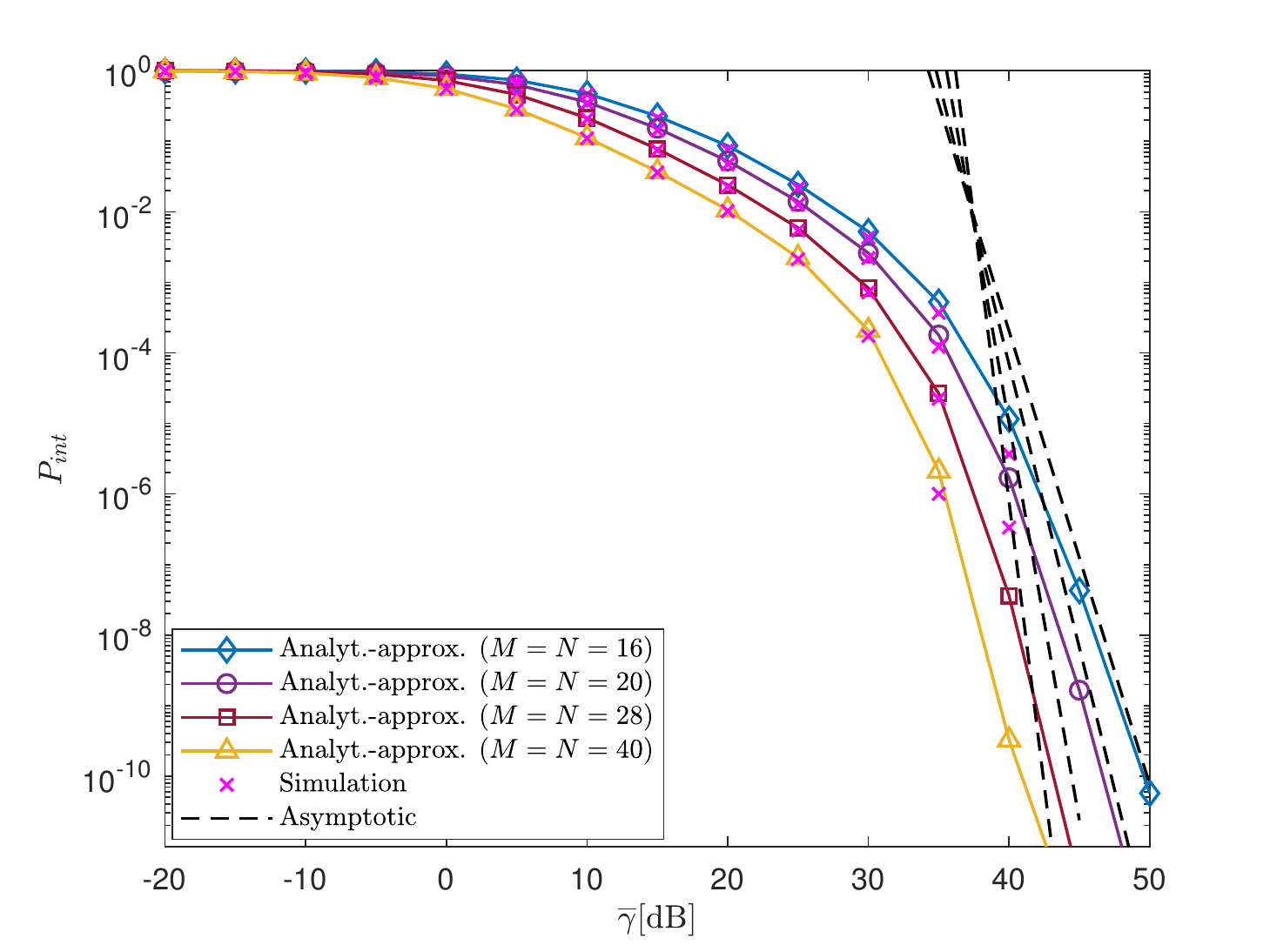}
\end{center}
\caption{IP versus $\overline{\protect\gamma }$ for various $(M,N)$ values.}
\label{fig2}
\end{figure}

In Fig. \ref{fig2}, the IP\ is plotted vs $\overline{\gamma }$ for various
values of REs' number $(M,N)$. Obviously, the system's secrecy improves by
increasing the number of REs, as pointed out in \textit{Remarks 1.2} and
\textit{3.2,} where one can attain an IP of $10^{-4}$ with $40$ REs for $%
\overline{\gamma }_{JE}=10$ dB even when the eavesdropper's link SNR\
exceeds the legitimate's ones by almost $10$ dB $\left( \text{i.e., }%
\overline{\gamma }=30\text{ dB, }\overline{\gamma }_{RE}=40\text{ dB}\right)
$. Also, the IP drops to $10^{-6}$ with $M,N\geq 28$ while $\overline{\gamma
}<\overline{\gamma }_{RE}$. Besides, it is noted that the high-SNR
asymptotic dashed curves, plotted from (\ref{ipass}), match tightly the
approximate's solid-line curves at high SNR values, corroborating the
accuracy of the asymptotic analysis.
\begin{figure}[tbp]
\begin{center}
\includegraphics[scale=.58]{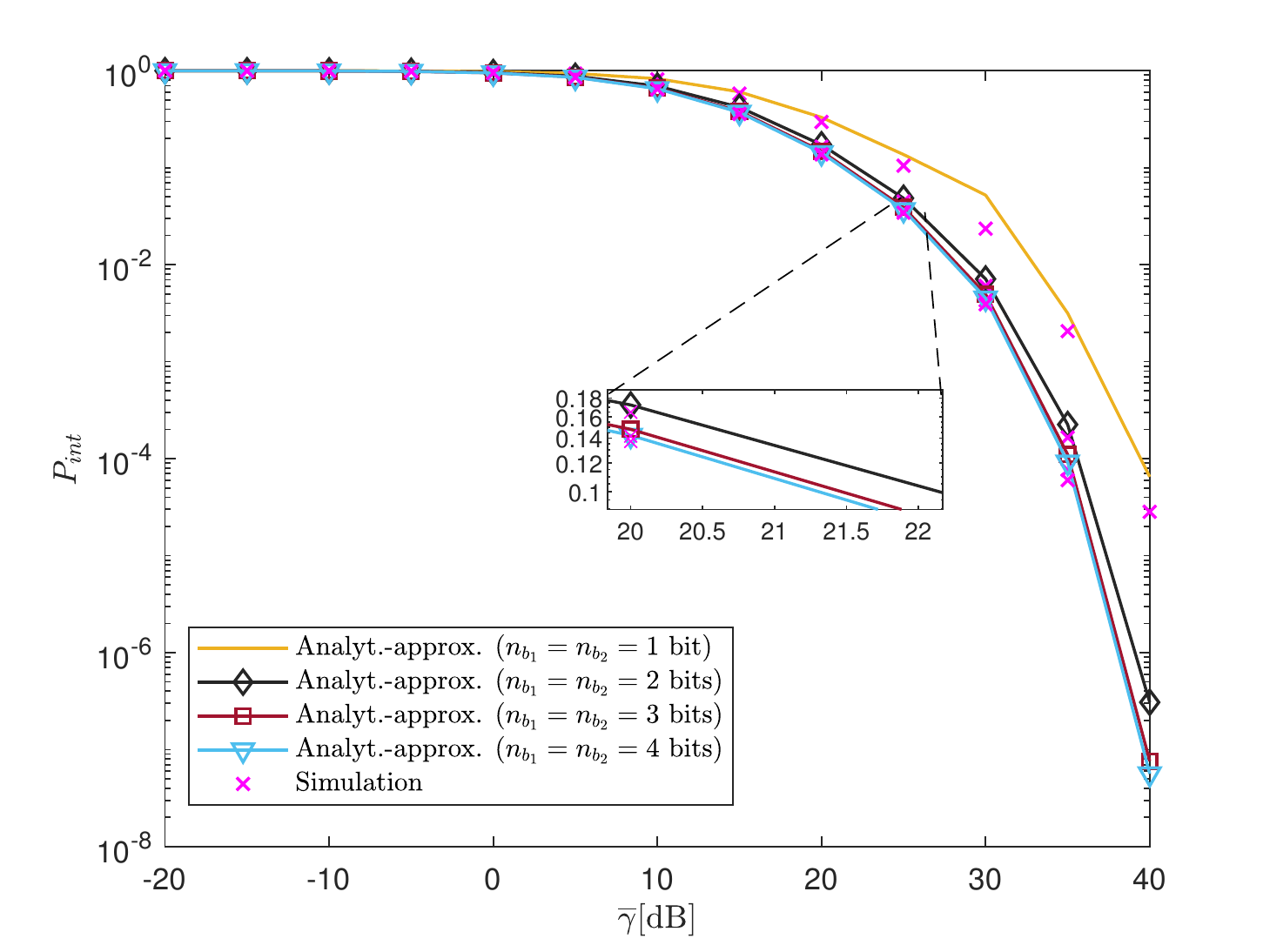}
\end{center}
\caption{IP versus $\overline{\protect\gamma }$ for various $\left(
n_{b_{1}},n_{b_{2}}\right) $ values.}
\label{fig3}
\end{figure}

The influence of the number of quantization bits (i.e., $n_{b_{1}},n_{b_{2}}$%
) is shown in Fig. \ref{fig3}, where the IP is plotted with respect to $%
\overline{\gamma }$ for various $n_{b_{1}}$ and $n_{b_{2}}$ values for $%
\overline{\gamma }_{JE}=0$ dB. One can ascertain that a very low phase
resolution, i.e., $n_{b_{1}}=n_{b_{2}}=1$ bit, yields a remarkable IP
degradation. Furthermore, the increase in $n_{b_{l}}$ $\left( l=1,2\right) $
to $2$ bits produces a $5$-dB secrecy gain at the high SNR, which endorses
the insights discussed in \textit{Remarks 1.1} and \textit{3.1}.
Importantly, we ascertain that the secrecy of the system is not improved
further when $n_{b_{l}}$ exceeds $3$ bits.

\begin{figure}[tbp]
\begin{center}
\includegraphics[scale=.58]{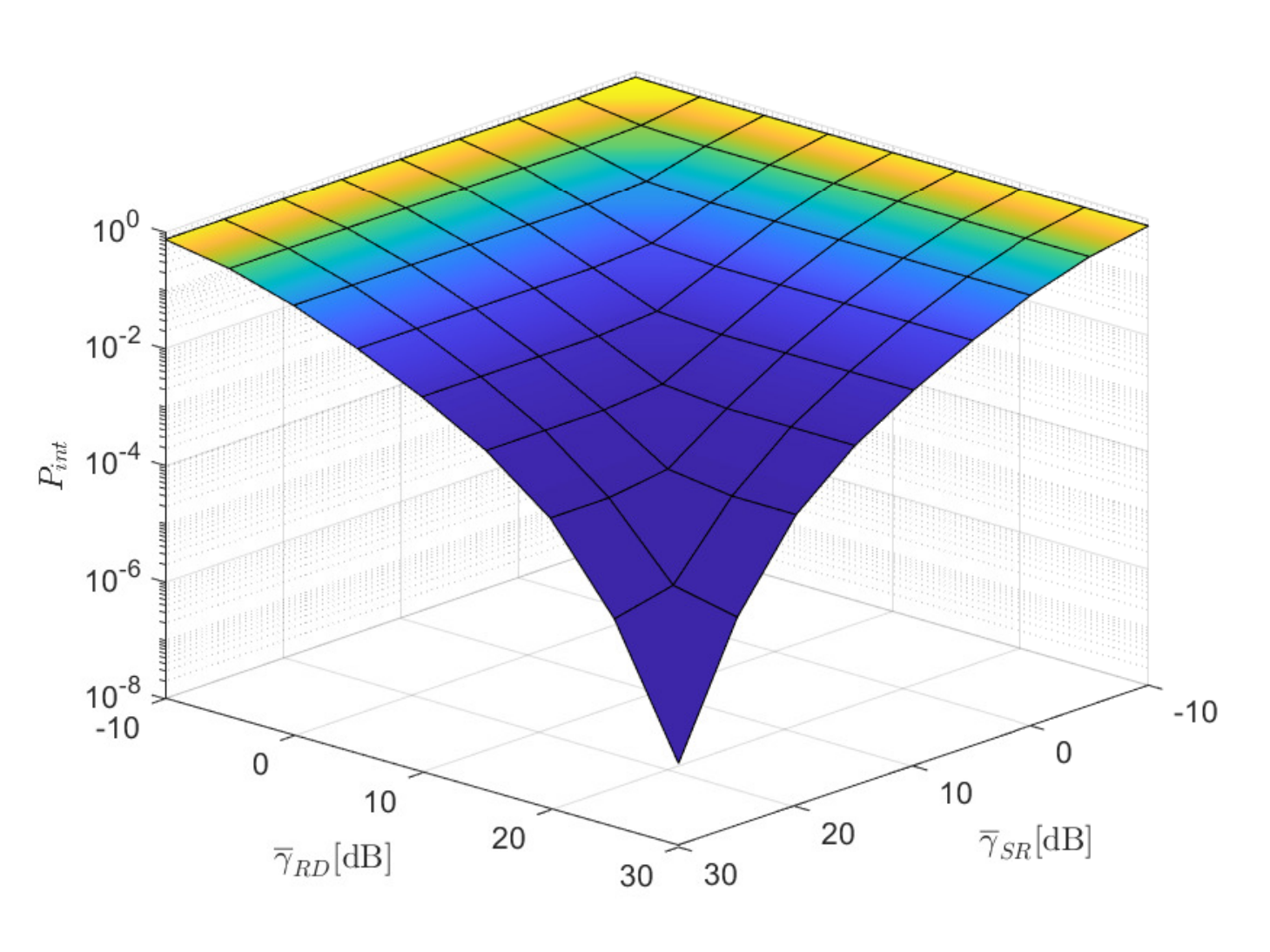}
\end{center}
\caption{IP versus $\overline{\protect\gamma }_{SR}$ and $\overline{\protect%
\gamma }_{RD}$.}
\label{fig4}
\end{figure}

The IP\ is shown in Fig. \ref{fig4} in three dimensions as a function of
both legitimate average SNRs ($\overline{\gamma }_{SR}$ and $\overline{%
\gamma }_{RD}$), where $\overline{\gamma }_{RE}=30$ dB and $M=N=24$. As
manifested in the previous figures, the IP\ decreases when both SNRs\
increase. Additionally, the IP\ exhibits a symmetric behavior for both
average SNRs, as discussed in \textit{Remark 3.3}. Furthermore, it is worth
noting that when $\overline{\gamma }_{SR}$ $\left( \overline{\gamma }%
_{RD}\right) $ is fixed, the IP reveals horizontal floors at high $\overline{%
\gamma }_{RD}$ $\left( \overline{\gamma }_{SR}\right) $. In fact, the e2e
SNR $\gamma _{eq}$, given by (\ref{gammaeq}), equals the minimum among $%
\gamma _{SR}$ and $\gamma _{RD}$. To this end, the IP\ reaches the
saturation regime at high $\overline{\gamma }_{SR}$ $\left( \overline{\gamma
}_{RD}\right) $ values when $\overline{\gamma }_{RD}$ $\left( \overline{%
\gamma }_{SR}\right) $ is fixed.
\begin{figure}[tbp]
\begin{center}
\includegraphics[scale=.58]{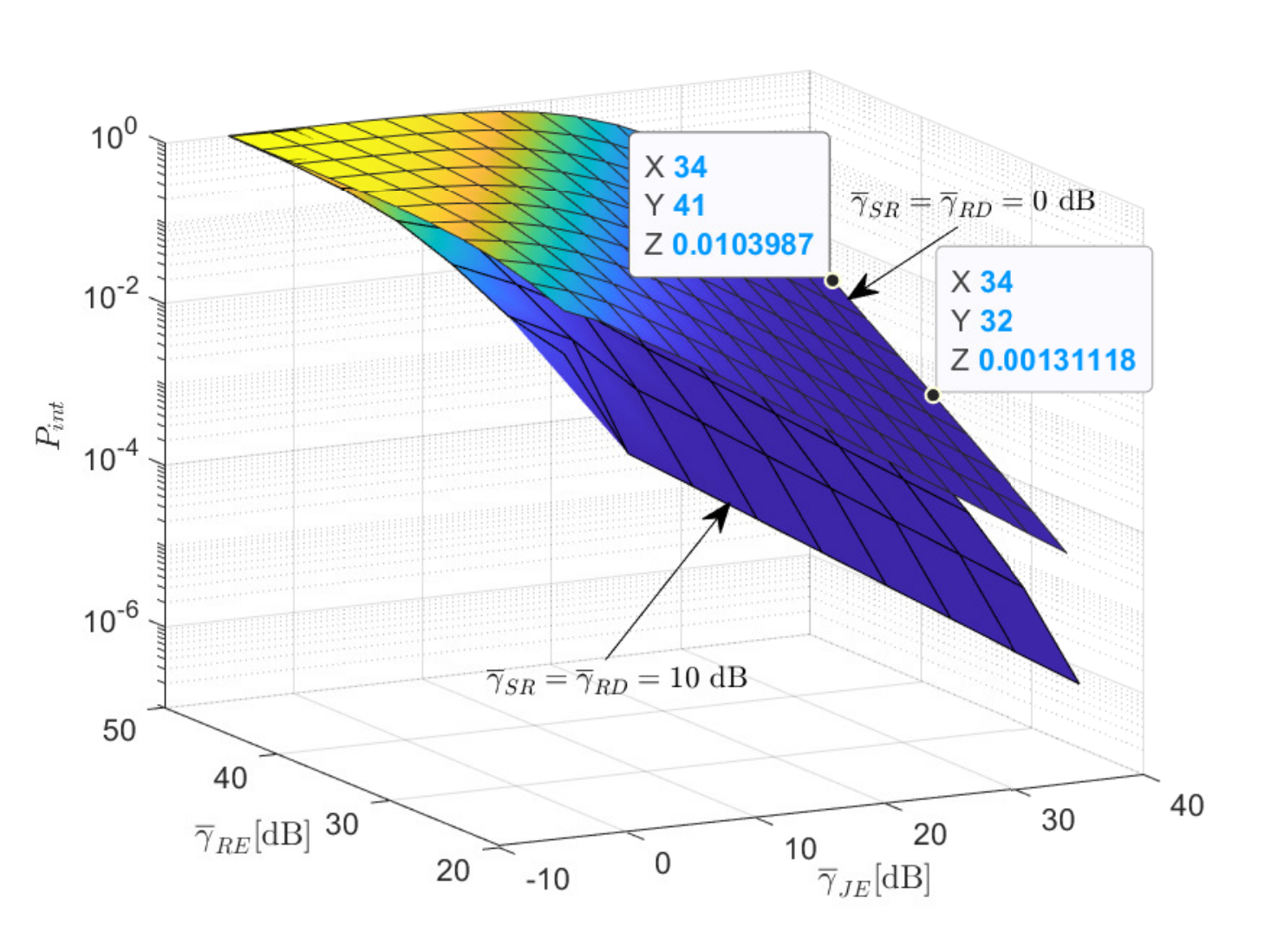}
\end{center}
\caption{IP versus $\overline{\protect\gamma }_{RE}$ and $\overline{\protect%
\gamma }_{JE}$.}
\label{fig5}
\end{figure}

Fig. \ref{fig5} presents the IP\ behavior versus $\overline{\gamma }_{RE}$
and $\overline{\gamma }_{JE}$, where $\overline{\gamma }_{SR}=\overline{%
\gamma }_{RD}=0,$ $10$ dB. Expectedly, the system's security level
deteriorates when $\overline{\gamma }_{RE}$ increases, i.e., $100\%$ IP is
manifested for $\overline{\gamma }_{SR}=\overline{\gamma }_{RD}=0$ dB, $%
\overline{\gamma }_{JE}\leq 0$ dB and $\overline{\gamma }_{RE}\geq 40$ dB.
Importantly, jamming can effectively improve the security where the IP\ can
be maintained around $10^{-2}$ and $1.3\times 10^{-3}$, for $\overline{\gamma
}_{JE}=34$ dB, even when the SNR\ of the illegitimate channel is $41$ and $%
32 $ dB-advantageous over the legitimate links', respectively.

\begin{figure}[tbp]
\begin{center}
\includegraphics[scale=.58]{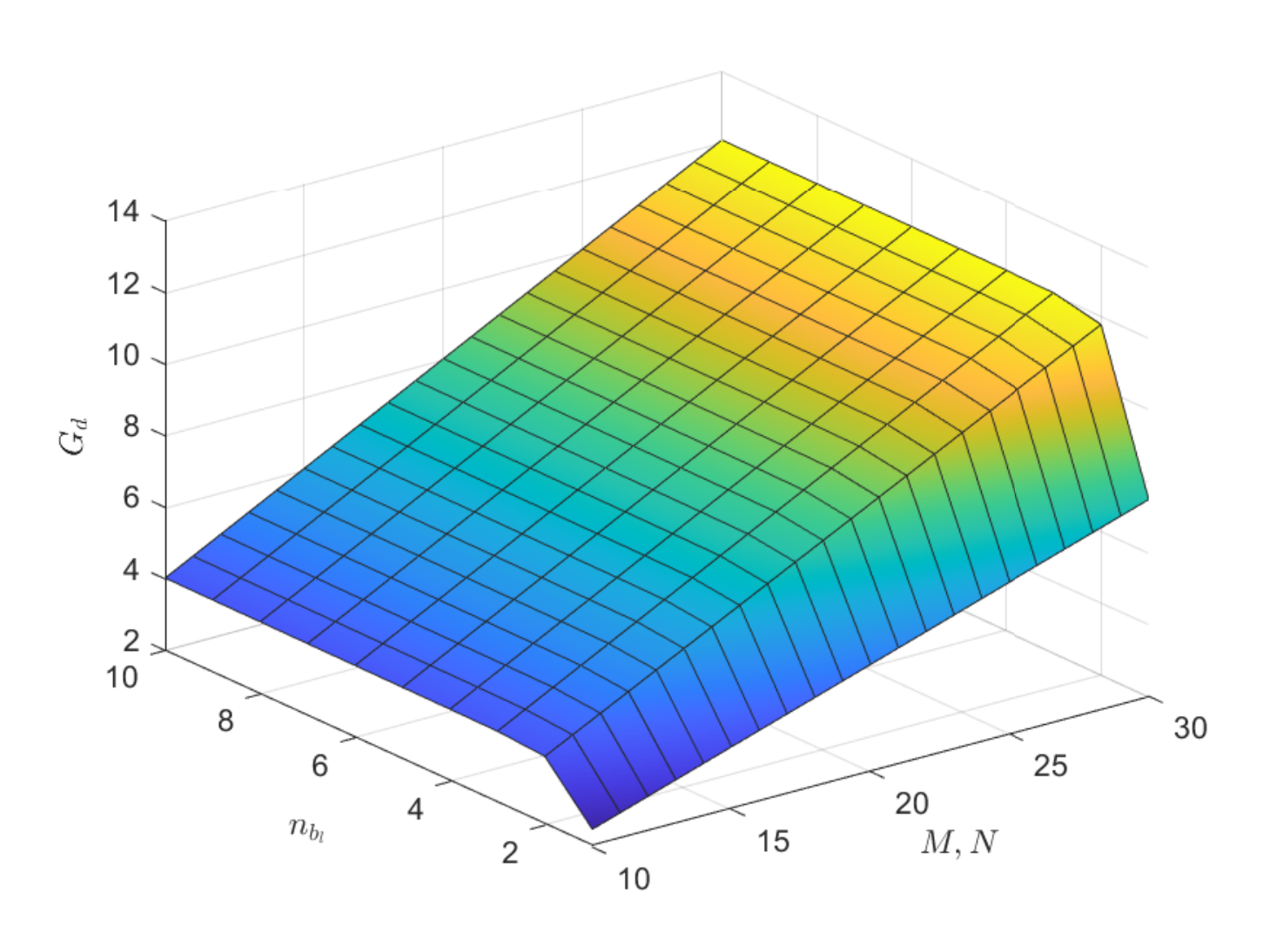}
\end{center}
\caption{Secrecy diversity order ($G_{d}$) versus $\left( M,N\right) $ and $%
n_{b_{l}}.$}
\label{fig7}
\end{figure}

\bigskip The system's secrecy diversity order is illustrated in Fig. \ref%
{fig7} in terms of the number of REs $(M,N)$ and $n_{b_{l}}$. As discussed
in \textit{Remark 4}, the system's diversity order increases with the
increase in the number of REs. That is, the higher the number of REs, the
greater the IP slope at high SNR as can be noted as well from the asymptotic
curves in Fig. \ref{fig2}. Also, the diversity order is less impacted by $%
n_{b_{l}}$ above $2$ bits.

\begin{figure}[tbp]
\begin{center}
\includegraphics[scale=.58]{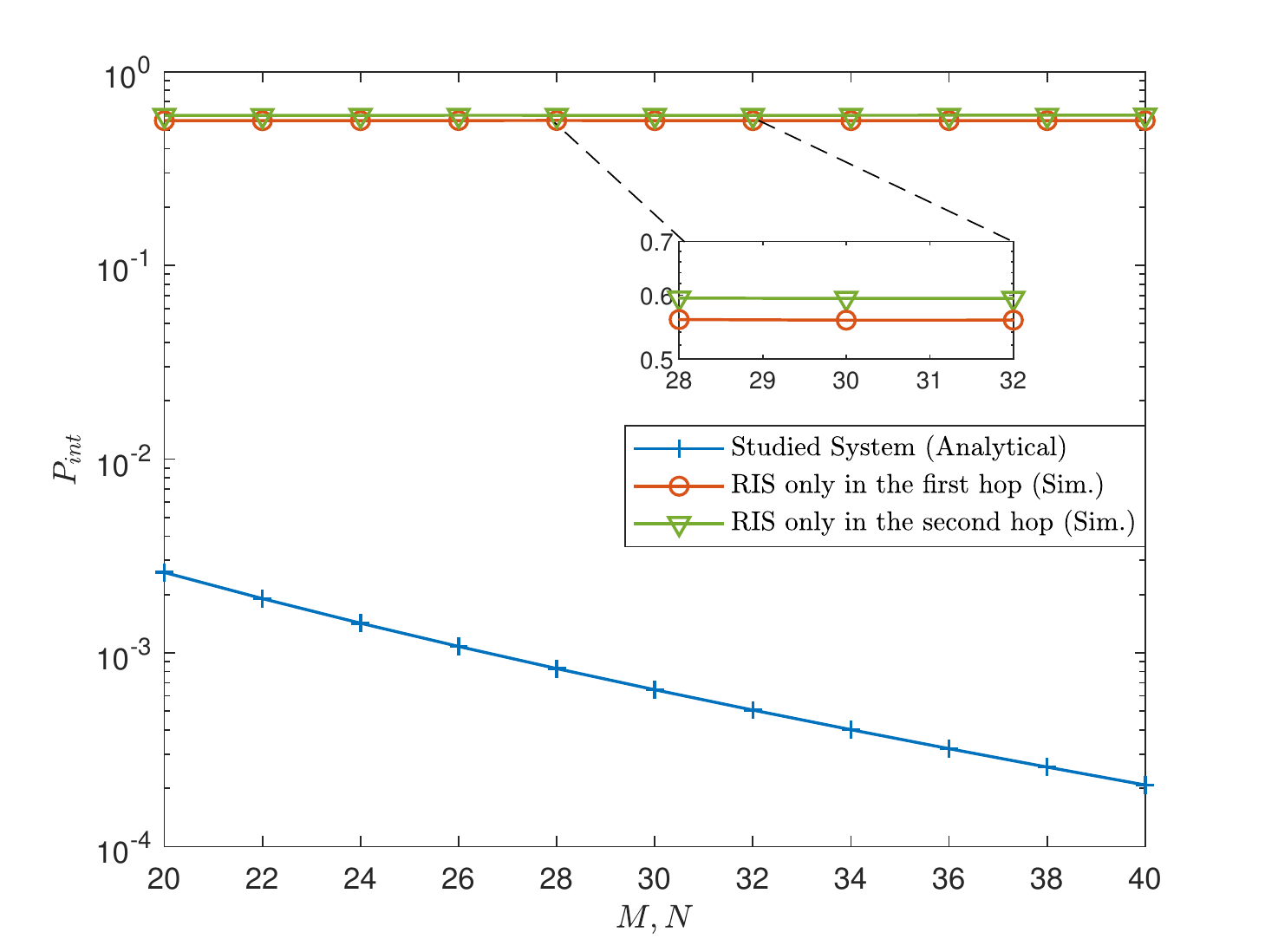}
\end{center}
\caption{IP\ versus $\overline{\protect\gamma }_{SR}$:\ A comparative
analysis with single-RIS dual-hop systems.}
\label{fig6}
\end{figure}

In Fig. \ref{fig6}, a comparative IP\ analysis is provided between the
considered system and two distinct setups of single-RIS\ dual-hop WCSs,
namely: when the RIS\ is incorporated only in the first hop (\textit{Setup I}%
) or the second hop (\textit{Setup II}). It is worthy to mention that the
results for \textit{Setups I}$\ $and \textit{II} were obtained by virtue of
Monte Carlo simulations. Also, we set $M=N$. The results show that the
analyzed system clearly outperforms the aforementioned schemes in terms of
PLS. Furthermore, we observe that \textit{Setup I}, i.e., RIS incorporated
only\ in the first hop, exhibits a slight secrecy improvement compared to
\textit{Setup II}, i.e., RIS\ in the second hop, where the eavesdropper and
legitimate destination in the latter case benefit from the SNR\ improvement
through involving the RIS in the second hop. Furthermore, it is obvious that
the analyzed system's IP manifests a decreasing behavior vs $(M,N)$,
different from\textit{\ Setups I} and \textit{II}, where the IP is constant
with respect to the number of REs. This is due to the saturation effect of
the DF relaying's e2e SNR, given by (\ref{gammaeq}), where $\gamma _{eq}$
equals $\gamma _{RD}$ regardless of boosting $\gamma _{SR}$\ by increasing
the number of REs (\textit{Setup I}) or vice-versa (\textit{Setup II}).

\section{Conclusion}

\label{concl}

In this paper, we provide a thorough analysis of the secrecy performance of
a dual-hop RIS-assisted and jamming-aided WCS. The setup consists of a
source transmitting information signals to a destination through a DF relay.
Two RISs assist the source-relay and relay-destination communications. We
assume the presence of an eavesdropper that attempts to overhear the signal
forwarded by the relay and reflected by the second RIS. We also introduce a
friendly jammer to increase the security level by means of a broadcasted
jamming signal. Finally, we assume that RISs phase shifts are subject to
PQEs.

Tight approximate and asymptotic IP\ expressions were derived, where the
impact of key system parameters was discussed. Our results showed that the
system's IP\ is significantly improved by increasing the jamming power and
number of REs, where the IP\ can reach $10^{-6}$ with $40$ REs and $10$ dB
of jamming power-to-noise ratio, even when the legitimate average SNRs are
below the wiretap one. Furthermore, we showed that there is no impact on the
secrecy of the system when the number of quantization bits exceeds $3$ bits.
Importantly, the results also illustrate the efficiency of jamming in
maintaining lower IP levels in strong eavesdropping scenarios (i.e., the
eavesdropper's SNR is much higher compared to the legitimate one) with a
fixed number of REs. We showed as well that the secrecy diversity order is
proportional to the minimum among both RISs' number of REs. Lastly, the
analyzed system shows remarkable secrecy improvement compared to the
single-RIS\ dual-hop networks, i.e., RIS involved in one of the two hops.

\section*{Acknowledgment}

This research was sponsored in part by the NATO Science for Peace and
Security Programme under grant SPS G5797.

\section*{Appendix A: Proof of Lemma 1}

First, let us compute the correlation coefficient between $g_{RE}$ and $%
g_{JE}$, given by (\ref{gre}) and (\ref{gje}), respectively, as follows%
\begin{equation}
\rho _{1}=\frac{\mathbb{E}\left[ g_{RE}g_{JE}\right] -\mu _{g_{RE}}\mu
_{g_{JE}}}{\sqrt{\sigma _{g_{RE}}\sigma _{g_{JE}}}},  \label{rho1}
\end{equation}%
where
\begin{eqnarray}
\mu _{g_{XY}} &=&\mathbb{E}\left[ g_{XY}\right] , \\
\sigma _{g_{XY}} &=&\mathbb{E}\left[ g_{XY}^{2}\right] -\mu _{g_{XY}}^{2}.
\end{eqnarray}%
with $XY\in \left\{ RD,RE,JE\right\} $. Leveraging (i)\ the linearity of the
expectation operator along with (ii) $\mathbb{E}\left[ h_{AB}\right] =0$ $%
\left( A,B\in \left\{ R,D,J,E,\left( L_{k}^{(2)}\right) _{1\leq k\leq
N}\right\} \right) $ for complex Gaussian distribution (i.e., Rayleigh
fading channels) and (iii) (\ref{phii}), (\ref{etai}), (\ref{omegai}), (\ref%
{gre}), and (\ref{gje}), one can find that $\mu _{g_{JE}}=\mu _{g_{RE}}=0$
regardless of the distributions of $\eta _{i}$ and $\varpi _{i}$. Thus, (\ref%
{rho1}) reduces to
\begin{equation}
\rho _{1}=\frac{\mathbb{E}\left[ g_{RE}g_{JE}\right] }{\sqrt{\sigma
_{g_{RE}}\sigma _{g_{JE}}}}.  \label{rho1new}
\end{equation}%
By plugging (\ref{gre}) and (\ref{gje}) into (\ref{rho1new}), its numerator
can be written as%
\begin{align}
\mathbb{E}\left[ g_{RE}g_{JE}\right] &
=\sum\limits_{i=1}^{N}\sum\limits_{k=1}^{N}\mathbb{E}\left[ \left\vert
h_{RL_{i}^{(2)}}\right\vert e^{j\left\{ \xi _{i}^{(2)}-\arg \left(
h_{L_{i}^{(2)}D}\right) \right\} }\right]  \notag \\
& \times \underset{=0}{\underbrace{\mathbb{E}\left[ h_{JL_{k}^{(2)}}\right] }%
}\mathbb{E}\left[ h_{L_{i}^{(2)}E}h_{L_{k}^{(2)}E}\right]  \notag \\
& \times \mathbb{E}\left[ e^{j\left\{ -\arg \left( h_{RL_{k}^{(2)}}\right)
-\arg \left( h_{L_{k}^{(2)}D}\right) +\xi _{k}^{(2)}\right\} }\right] ,
\end{align}%
which vanishes to $0$. Therefore, $g_{RE}$ and $g_{JE}$ are uncorrelated. By
following a similar rationale for the correlations between ($g_{RD}$ and $%
g_{RE}$)\ and ($g_{RD}$ and $g_{JE}$), one can find that the respective
correlation coefficients are sums of products involving the expectation of $%
h_{L_{k}^{(2)}E}$ multiplied by the average of the remaining involved links'
fading envelopes, phase shifts, and PQEs. Thus, as $\mathbb{E}\left[ h_{AB}%
\right] =0$, both correlation coefficients vanish, resulting in the mutual
independence of $g_{RD}$, $g_{RE}$, and $g_{JE}$. This concludes the proof
of \textit{Lemma 1}.

\section*{Appendix B: Proof of Proposition 1}

From the IP\ definition in (\ref{ipdef}), we can express it as follows
\begin{equation}
P_{int}=1-\int_{0}^{\infty }F_{\gamma _{eq}}^{(c)}\left( x\right) f_{\gamma
_{RE}}\left( x\right) dx.  \label{ip1}
\end{equation}

Thus, we attain (\ref{ipb}) shown at the top of the next page,
\begin{figure*}[t]
{\normalsize 
\setcounter{mytempeqncnt}{\value{equation}}
\setcounter{equation}{54} }
\par
\begin{align}
P_{int}& \overset{(a)}{\approx }1-\frac{1}{N\Gamma \left( m_{SR}\right)
\Gamma \left( m_{RD}\right) }\int_{0}^{\infty }G_{1,2}^{2,0}\left( \frac{%
m_{SR}}{\Omega _{SR}\overline{\mathcal{\gamma }}_{SR}}x\left\vert
\begin{array}{c}
-;1 \\
0,m_{SR}%
\end{array}%
\right. \right) G_{1,2}^{2,0}\left( \frac{m_{RD}}{\Omega _{RD}\overline{%
\gamma }_{RD}}x\left\vert
\begin{array}{c}
-;1 \\
0,m_{RD}%
\end{array}%
\right. \right)  \notag \\
& \times \frac{\exp \left( -\frac{x}{N\overline{\mathcal{\gamma }}_{RE}}%
\right) \left[ \overline{\mathcal{\gamma }}_{JE}N\overline{\mathcal{\gamma }}%
_{RE}+\overline{\mathcal{\gamma }}_{RE}+\overline{\mathcal{\gamma }}_{JE}x%
\right] }{\left( \overline{\mathcal{\gamma }}_{RE}+\overline{\mathcal{\gamma
}}_{JE}x\right) ^{2}}dx,  \notag \\
& \overset{(b)}{=}1-\frac{1}{\left( 2\pi j\right) ^{2}N\Gamma \left(
m_{SR}\right) \Gamma \left( m_{RD}\right) }\int_{C_{s}}\int_{C_{v}}\frac{%
\Gamma \left( s\right) \Gamma \left( m_{SR}+s\right) \Gamma \left( v\right)
\Gamma \left( m_{RD}+v\right) }{\Gamma \left( 1+s\right) \Gamma \left(
1+v\right) }\int_{0}^{\infty }\left[ \overline{\gamma }_{JE}\left( N%
\overline{\gamma }_{RE}+x\right) +\overline{\gamma }_{RE}\right]  \notag \\
& \times \underset{\triangleq F(s,v)=F_{1}(s,v)+F_{2}(s,v)}{\underbrace{%
\frac{\exp \left( -\frac{x}{N\overline{\gamma }_{RE}}\right) }{\left(
\overline{\gamma }_{RE}+\overline{\gamma }_{JE}x\right) ^{2}}x^{-s-v}dx}}%
\left( \frac{m_{SR}}{\Omega _{SR}\overline{\gamma }_{SR}}\right) ^{-s}\left(
\frac{m_{RD}}{\Omega _{RD}\overline{\gamma }_{RD}}\right) ^{-v}dsdv,
\label{ipb}
\end{align}%
\par
{\normalsize 
\hrulefill 
\vspace*{4pt} }
\end{figure*}
where \textit{Step (a)} is obtained by involving the CCDF and PDF\
expressions from (\ref{ccdfeq}) and (\ref{pdf}) into (\ref{ip1}) along with
using \cite[Eq. (06.06.26.0005.01)]{wolfram}, while \textit{Step (b) }is
produced via the Mellin-Barnes integral (MBI)\ definition \cite[Eq. (1.112)]%
{mathai} with%
\begin{align}
\mathcal{F}_{i}(s,v)& \triangleq \left( \overline{\mathcal{\gamma }}_{JE}N%
\overline{\mathcal{\gamma }}_{RE}\right) ^{i-1}\int_{0}^{\infty }\frac{\exp
\left( -\frac{x}{N\overline{\mathcal{\gamma }}_{RE}}\right) }{\left(
\overline{\mathcal{\gamma }}_{RE}+\overline{\mathcal{\gamma }}_{JE}x\right)
^{i}}x^{-s-v}dx.  \notag \\
& \overset{(a)}{=}\left( \overline{\mathcal{\gamma }}_{JE}N\overline{%
\mathcal{\gamma }}_{RE}\right) ^{i-1}\int_{0}^{\infty }\frac{x^{-s-v}}{%
\left( \overline{\mathcal{\gamma }}_{RE}+\overline{\mathcal{\gamma }}%
_{JE}x\right) ^{i}}  \notag \\
& \times G_{0,1}^{1,0}\left( \frac{x}{N\overline{\mathcal{\gamma }}_{RE}}%
\left\vert
\begin{array}{c}
-;- \\
0;-%
\end{array}%
\right. \right) dx,  \notag \\
& \overset{(b)}{=}\frac{\left( \overline{\mathcal{\gamma }}_{JE}N\overline{%
\mathcal{\gamma }}_{RE}\right) ^{i-1}}{2\pi j}\int_{C_{w}}\Gamma \left(
w\right) \left( \frac{1}{N\overline{\mathcal{\gamma }}_{RE}}\right) ^{-w}
\notag \\
& \times \int_{0}^{\infty }\frac{x^{-s-v-w}}{\left( \overline{\mathcal{%
\gamma }}_{RE}+\overline{\mathcal{\gamma }}_{JE}x\right) ^{i}}dxdw,  \notag
\\
& \overset{(c)}{=}\frac{\overline{\mathcal{\gamma }}_{JE}^{i-2}N^{i-1}}{2\pi
j}\int_{C_{w}}\Gamma \left( w\right) \left( \frac{1}{N\overline{\mathcal{%
\gamma }}_{JE}}\right) ^{-w}\left( \frac{\overline{\mathcal{\gamma }}_{RE}}{%
\overline{\mathcal{\gamma }}_{JE}}\right) ^{-s}  \notag \\
& \times \Gamma \left( 1-s-v-w\right) \Gamma \left( i-1+s+v+w\right) dw.
\label{Ic}
\end{align}%
\textit{Step (a)} of (\ref{Ic}) yields by means of \cite[Eq.
(07.34.03.0228.01)]{wolfram}, while using \cite[Eq. (1.112)]{mathai}
produces \textit{Step (b)}. Finally, leveraging \cite[Eqs. (3.194.3, 8.384.1)%
]{integrals} , \textit{Step (c)}\ holds.

By plugging (\ref{Ic}) into (\ref{ipb}), it yields (\ref{ippenult}) at the
top of the page.
\begin{figure*}[t]
{\normalsize 
\setcounter{mytempeqncnt}{\value{equation}}
\setcounter{equation}{56} }
\par
\begin{align}
P_{int}& \approx 1-\frac{1}{\left( 2\pi j\right) ^{3}\Gamma \left(
m_{SR}\right) \Gamma \left( m_{RD}\right) }\sum\limits_{i=1}^{2}\left(
\overline{\mathcal{\gamma }}_{JE}N\right)
^{i-2}\int_{C_{s}}\int_{C_{v}}\int_{C_{w}}\frac{\Gamma \left( s\right)
\Gamma \left( m_{SR}+s\right) \Gamma \left( v\right) \Gamma \left(
m_{RD}+v\right) }{\Gamma \left( 1+s\right) \Gamma \left( 1+v\right) }  \notag
\\
& \times \Gamma \left( w\right) \Gamma \left( 1-s-v-w\right) \Gamma \left(
i-1+s+v+w\right) \left( \frac{m_{SR}}{\Omega _{SR}\overline{\mathcal{\gamma }%
}_{SR}}\frac{\overline{\mathcal{\gamma }}_{RE}}{\overline{\mathcal{\gamma }}%
_{JE}}\right) ^{-s}\left( \frac{m_{RD}}{\Omega _{RD}\overline{\mathcal{%
\gamma }}_{RD}}\frac{\overline{\mathcal{\gamma }}_{RE}}{\overline{\mathcal{%
\gamma }}_{JE}}\right) ^{-v}\left( \frac{1}{N\overline{\mathcal{\gamma }}%
_{JE}}\right) ^{-w}dsdvdw.  \label{ippenult}
\end{align}%
\par
{\normalsize 
\hrulefill 
\vspace*{4pt} }
\end{figure*}

Importantly, armed by the residues theorem, such a triple MBI can be reduced
to a double\ MBI by evaluating the series of residues on the left half-plane
poles of both $\Gamma \left( w\right) $ and $\Gamma \left( i-1+s+v+w\right) $
as shown in (\ref{doublembi}) at the top of the next page \cite[Theorem 1.2]%
{kilbas}
\begin{figure*}[t]
{\normalsize 
\setcounter{mytempeqncnt}{\value{equation}}
\setcounter{equation}{57} }
\par
\begin{align}
P_{int}& \approx 1-\frac{1}{\Gamma \left( m_{SR}\right) \Gamma \left(
m_{RD}\right) \left( 2\pi j\right) ^{2}}\sum\limits_{l=0}^{\infty }\frac{%
\left( -1\right) ^{l}}{l!}\sum\limits_{i=1}^{2}\left( \overline{\mathcal{%
\gamma }}_{JE}N\right) ^{i-2}\int_{C_{s}}\int_{C_{v}}\frac{\Gamma \left(
s\right) \Gamma \left( m_{SR}+s\right) \Gamma \left( v\right) \Gamma \left(
m_{RD}+v\right) }{\Gamma \left( 1+s\right) \Gamma \left( 1+v\right) }  \notag
\\
& \times \left( \frac{m_{SR}}{\Omega _{SR}\overline{\mathcal{\gamma }}_{SR}}%
\frac{\overline{\mathcal{\gamma }}_{RE}}{\overline{\mathcal{\gamma }}_{JE}}%
\right) ^{-s}\left( \frac{m_{RD}}{\Omega _{RD}\overline{\mathcal{\gamma }}%
_{RD}}\frac{\overline{\mathcal{\gamma }}_{RE}}{\overline{\mathcal{\gamma }}%
_{JE}}\right) ^{-v}\left[
\begin{array}{c}
\frac{\Gamma \left( 1-s-v+l\right) \Gamma \left( i-1+s+v-l\right) }{\left( N%
\overline{\mathcal{\gamma }}_{JE}\right) ^{l}} \\
+\frac{\Gamma \left( -s-v-l-i+1\right) \Gamma \left( l+i\right) }{\left( N%
\overline{\mathcal{\gamma }}_{JE}\right) ^{s+v+l+i-1}}%
\end{array}%
\right] dsdv.  \label{doublembi}
\end{align}%
\par
{\normalsize 
\hrulefill 
\vspace*{4pt} }
\end{figure*}

Finally, relying on the bivariate Fox's $H$-function definition in \cite[%
Eqs. (1.11-1.13)]{yakub} along with some algebraic manipulations and
simplifications, (\ref{ipfinal}) is reached. This concludes the
proposition's proof.

%

\section*{Appendix C:\ Proof of Proposition 2}

At the high SNR\ regime ($\overline{\gamma }_{SR}=\overline{\gamma }_{RD}=%
\overline{\gamma }\rightarrow \infty $), the IP can be expanded as follows

\begin{align}
P_{int}^{\left( \infty \right) }& \sim 1-\int_{0}^{\infty }F_{\gamma
_{eq}}^{\left( c,\infty \right) }\left( x\right) f_{\gamma _{RE}}\left(
x\right) dx  \label{ipasdef} \\
& \overset{(a)}{=}1-\frac{1}{N}\int_{0}^{\infty }\frac{\left[ \overline{%
\mathcal{\gamma }}_{JE}N\overline{\mathcal{\gamma }}_{RE}+\overline{\mathcal{%
\gamma }}_{RE}+\overline{\mathcal{\gamma }}_{JE}x\right] }{\left( \overline{%
\mathcal{\gamma }}_{RE}+\overline{\mathcal{\gamma }}_{JE}x\right) ^{2}}
\notag \\
& \times \exp \left( -\frac{x}{N\overline{\mathcal{\gamma }}_{RE}}\right)
\frac{\left( \Gamma \left( m_{SR}\right) -\frac{\left( \frac{m_{SR}}{\Omega
_{SR}\overline{\mathcal{\gamma }}}x\right) ^{m_{SR}}}{m_{SR}}\right) }{%
\Gamma \left( m_{SR}\right) }  \notag \\
& \times \frac{\left( \Gamma \left( m_{RD}\right) -\frac{\left( \frac{m_{RD}%
}{\Omega _{RD}\overline{\mathcal{\gamma }}}x\right) ^{m_{RD}}}{m_{RD}}%
\right) }{\Gamma \left( m_{RD}\right) }dx  \label{ipasdefstpa} \\
& \overset{(b)}{=}\mathcal{G}_{1}+\mathcal{G}_{2}-\mathcal{G}_{3},
\label{ipas}
\end{align}%
with $\left( \mathcal{G}_{i}\right) 1\leq i\leq 3$ are given by (\ref{G1})-(%
\ref{G3}) in the next page, where \textit{Step (a)} in (\ref{ipasdefstpa})
is reached by incorporating the CCDF\ of $\gamma _{eq}$ in (\ref{ccdfeq})
and the PDF of $\gamma _{RE}$ in (\ref{pdf}) into (\ref{ipasdef}) along with
the upper incomplete Gamma expansion \cite[Eq. (06.06.06.0001.02)]{wolfram},
while \textit{Step (b) }yields\textit{\ }by expanding \textit{Step (a) }and
taking into account that $\int_{0}^{\infty }f_{\gamma _{RE}}\left( x\right)
dx=1$.
\begin{figure*}[t]
{\normalsize 
\setcounter{mytempeqncnt}{\value{equation}}
\setcounter{equation}{61} }
\par
\begin{eqnarray}
\mathcal{G}_{1} &=&\frac{\overline{\mathcal{\gamma }}^{-m_{SR}}}{\Gamma
\left( m_{SR}\right) }\int_{0}^{\infty }\frac{\left( \frac{m_{SR}}{\Omega
_{SR}}x\right) ^{m_{SR}}}{m_{SR}}\frac{\exp \left( -\frac{x}{N\overline{%
\mathcal{\gamma }}_{RE}}\right) \left[ \overline{\mathcal{\gamma }}_{JE}N%
\overline{\mathcal{\gamma }}_{RE}+\overline{\mathcal{\gamma }}_{RE}+%
\overline{\mathcal{\gamma }}_{JE}x\right] }{N\left( \overline{\mathcal{%
\gamma }}_{RE}+\overline{\mathcal{\gamma }}_{JE}x\right) ^{2}}dx.  \label{G1}
\\
\mathcal{G}_{2} &=&\frac{\overline{\mathcal{\gamma }}^{-m_{RD}}}{\Gamma
\left( m_{RD}\right) }\int_{0}^{\infty }\frac{\left( \frac{m_{RD}}{\Omega
_{RD}}x\right) ^{m_{RD}}}{m_{RD}}\frac{\exp \left( -\frac{x}{N\overline{%
\mathcal{\gamma }}_{RE}}\right) \left[ \overline{\mathcal{\gamma }}_{JE}N%
\overline{\mathcal{\gamma }}_{RE}+\overline{\mathcal{\gamma }}_{RE}+%
\overline{\mathcal{\gamma }}_{JE}x\right] }{N\left( \overline{\mathcal{%
\gamma }}_{RE}+\overline{\mathcal{\gamma }}_{JE}x\right) ^{2}}dx.  \label{G2}
\\
\mathcal{G}_{3} &=&\frac{\overline{\mathcal{\gamma }}^{-\left(
m_{SR}+m_{RD}\right) }}{\Gamma \left( m_{SR}\right) \Gamma \left(
m_{RD}\right) }\int_{0}^{\infty }\frac{\left( \frac{m_{SR}}{\Omega _{SR}}%
x\right) ^{m_{SR}}}{m_{SR}}\frac{\left( \frac{m_{RD}}{\Omega _{RD}}x\right)
^{m_{RD}}}{m_{RD}}\frac{\exp \left( -\frac{x}{N\overline{\mathcal{\gamma }}%
_{RE}}\right) \left[ \overline{\mathcal{\gamma }}_{JE}N\overline{\mathcal{%
\gamma }}_{RE}+\overline{\mathcal{\gamma }}_{RE}+\overline{\mathcal{\gamma }}%
_{JE}x\right] }{N\left( \overline{\mathcal{\gamma }}_{RE}+\overline{\mathcal{%
\gamma }}_{JE}x\right) ^{2}}dx.  \label{G3}
\end{eqnarray}%
\par
{\normalsize 
\hrulefill 
\vspace*{4pt} }
\end{figure*}

\bigskip Importantly, one can note that the IP\ expression in (\ref{ipas})
is the sum of three terms, given by (\ref{G1})-(\ref{G3}), where each of
which has a different coding gain and diversity order pair. Thus, the IP\
will be expanded by the term having the lowest power of $\overline{\mathcal{%
\gamma }}$. Henceforth, three different cases are distinguished, namely:

\subsection{Case I:\ $M<N$}

Since $m_{SR}$ and $m_{RD}$ are proportional to $M$ and $N$, respectively,
the IP\ will be expanded by $\mathcal{G}_{1}$ in (\ref{G1})\ for which the
diversity order equals $m_{SR}$. Henceforth, it yields (\ref{ipas1}) at the
top of the page, where \textit{Step (a)}\ holds by decomposing (\ref{G1}%
) into two terms, while \textit{Step (b)}\ is reached via the use of
integration by parts with $f^{\prime }(x)=\left( \overline{\mathcal{\gamma }}%
_{RE}+\overline{\mathcal{\gamma }}_{JE}x\right) ^{-2}$ and $%
g(x)=x^{m_{SR}}\exp \left( -\frac{x}{N\overline{\mathcal{\gamma }}_{RE}}%
\right) $. Finally, by utilizing \cite[Eq. (3.383.10)]{integrals} along with
some algebraic manipulations, the first case of (\ref{Gc}) is attained.
\begin{figure*}[t]
{\normalsize 
\setcounter{mytempeqncnt}{\value{equation}}
\setcounter{equation}{64} }
\par
\begin{eqnarray}
&&\mathcal{G}_{1}\overset{(a)}{=}\overline{\mathcal{\gamma }}^{-m_{SR}}%
\underset{\triangleq G_{c}}{\underbrace{\frac{\left( \frac{m_{SR}}{\Omega
_{SR}}\right) ^{m_{SR}}}{N\Gamma \left( m_{SR}+1\right) }\left[ \frac{1}{%
\overline{\mathcal{\gamma }}_{JE}}\int_{0}^{\infty }x^{m_{SR}}\frac{\exp
\left( -\frac{x}{N\overline{\mathcal{\gamma }}_{RE}}\right) }{\frac{%
\overline{\mathcal{\gamma }}_{RE}}{\overline{\mathcal{\gamma }}_{JE}}+x}dx+%
\overline{\mathcal{\gamma }}_{JE}N\overline{\mathcal{\gamma }}%
_{RE}\int_{0}^{\infty }x^{m_{SR}}\frac{\exp \left( -\frac{x}{N\overline{%
\mathcal{\gamma }}_{RE}}\right) }{\left( \overline{\mathcal{\gamma }}_{RE}+%
\overline{\mathcal{\gamma }}_{JE}x\right) ^{2}}dx\right] }}  \notag \\
&&\overset{(b)}{=}\overline{\mathcal{\gamma }}^{-m_{SR}}\frac{\left( \frac{%
m_{SR}}{\Omega _{SR}}\right) ^{m_{SR}}}{N\Gamma \left( m_{SR}+1\right) }%
\left[
\begin{array}{c}
\frac{1}{\overline{\mathcal{\gamma }}_{JE}}\int_{0}^{\infty }x^{m_{SR}}\frac{%
\exp \left( -\frac{x}{N\overline{\mathcal{\gamma }}_{RE}}\right) }{\frac{%
\overline{\mathcal{\gamma }}_{RE}}{\overline{\mathcal{\gamma }}_{JE}}+x}dx-N%
\overline{\mathcal{\gamma }}_{RE}\underset{=0}{\underbrace{\left[ \frac{%
x^{m_{SR}}\exp \left( -\frac{x}{N\overline{\mathcal{\gamma }}_{RE}}\right) }{%
\overline{\mathcal{\gamma }}_{RE}+\overline{\mathcal{\gamma }}_{JE}x}\right]
_{0}^{\infty }}} \\
+N\overline{\mathcal{\gamma }}_{RE}\int_{0}^{\infty }\frac{\left[
m_{SR}x^{m_{SR}-1}\exp \left( -\frac{x}{N\overline{\mathcal{\gamma }}_{RE}}%
\right) -\frac{1}{N\overline{\mathcal{\gamma }}_{RE}}x^{m_{SR}}\exp \left( -%
\frac{x}{N\overline{\mathcal{\gamma }}_{RE}}\right) \right] }{\overline{%
\mathcal{\gamma }}_{RE}+\overline{\mathcal{\gamma }}_{JE}x}dx%
\end{array}%
\right] .  \label{ipas1}
\end{eqnarray}%
\par
{\normalsize 
\hrulefill 
\vspace*{4pt} }
\end{figure*}

\subsection{Case II:\ $M>N$}

\bigskip In this case, the IP\ will be expanded by $\mathcal{G}_{2}$. By
following a similar rationale to \textit{Case I} and substituting the index $%
SR$ by $RD$, we reach the second case of (\ref{Gc}).

\subsection{Case III:\ $M=N$}

In this case, the IP\ is expanded by the sum of the two terms produced in
the previous cases, yielding the third case of (\ref{Gc}). This concludes
the proof of \textit{Proposition 2}.

\bibliographystyle{IEEEtran}
\bibliography{refs}

\end{document}